\documentclass[fleqn,usenatbib]{mnras}

\usepackage{newtxtext,newtxmath}
\usepackage[T1]{fontenc}
\usepackage{ae,aecompl}

\usepackage{graphicx}	% Including figure files
\usepackage{amsmath}	% Advanced maths commands
\usepackage{amssymb}	% Extra maths symbols
\usepackage{url}

\def\swift{{\it Swift}}
\newcommand{\nh}{\hbox{$N_{\rm H}$}~}
\newcommand{\hcm}[1]{$\times 10^{#1}$ cm$^{-2}$}
\newcommand{\ergs}[1]{$\times 10^{#1}$ \hbox{erg s$^{-1}$}}

\newcommand{\cts}[1]{$\times 10^{#1}$ ct s$^{-1}$}

\newcommand{\tpower}[1]{$\times 10^{#1}$}
\newcommand{\power}[1]{$10^{#1}$}
\newcommand{\appropto}{\mathrel{\vcenter{
  \offinterlineskip\halign{\hfil$##$\cr
    \propto\cr\noalign{\kern1pt}\sim\cr\noalign{\kern-2pt}}}}}

\title[The 2015 Nova Eruption in IC\,1613]{Multi-wavelength Observations of the 2015 Nova in the Local Group Irregular Dwarf Galaxy IC\,1613}

\author[S.~C. Williams et al.]{
S.~C. Williams,$^{1,2}$\thanks{E-mail: s.williams7@lancaster.ac.uk}
M.~J. Darnley,$^{2}$\thanks{E-mail: M.J.Darnley@ljmu.ac.uk}
and M. Henze$^{3}$\thanks{E-mail: henze@ice.cat}
\\
$^{1}$Physics Department, Lancaster University, Lancaster, LA1 4YB, UK\\
$^{2}$Astrophysics Research Institute, Liverpool John Moores University, IC2 Liverpool Science Park, Liverpool, L3 5RF, UK\\
$^{3}$Institute of Space Sciences (IEEC-CSIC), Campus UAB, Carrer de Can Magrans, s/n 08193 Barcelona, Spain
}

\date{Accepted XXX. Received YYY; in original form ZZZ}

\pubyear{2017}

\begin{document}
\label{firstpage}
\pagerange{\pageref{firstpage}--\pageref{lastpage}}
\maketitle

% Abstract of the paper
\begin{abstract}
A nova in the Local Group irregular dwarf galaxy IC\,1613 was discovered on 2015 September 10 and is the first nova in that galaxy to be spectroscopically confirmed. We conducted a detailed multi-wavelength observing campaign of the eruption with the Liverpool Telescope, the LCO 2\,m telescope at Siding Spring Observatory, and {\it Swift}, the results of which we present here. The nova peaked at $M_V=-7.93\pm0.08$ and was fast-fading, with decline times of $t_{2(V)}=13\pm2$ and $t_{3(V)}=26\pm2$\,days. The overall light curve decline was relatively smooth, as often seen in fast-fading novae. {\it Swift} observations spanned 40\,days to 332\,days post-discovery, but no X-ray source was detected. Optical spectra show the nova to be a member of the hybrid spectroscopic class, simultaneously showing Fe\,{\sc ii} and N\,{\sc ii} lines of similar strength during the early decline phase. The spectra cover the eruption from the early optically thick phase, through the early decline and into the nebular phase. The H$\gamma$ absorption minimum from the optically thick spectrum indicates an expansion velocity of $1200\pm200$\,km\,s$^{-1}$. The FWHM of the H$\alpha$ emission line between 10.54 and 57.51\,days post-discovery shows no significant evolution and remains at $\sim1750$\,km\,s$^{-1}$, although the morphology of this line does show some evolution. The nova appears close to a faint stellar source in archival imaging, however we find the most likely explanation for this is simply a chance alignment.
\end{abstract}

% Select between one and six entries from the list of approved keywords.
% Don't make up new ones.

%%%%%%%%%%%%%%%%%%%%%%%%%%%%%%%%%%%%%%%%%%%%%%%%%%

%%%%%%%%%%%%%%%%% BODY OF PAPER %%%%%%%%%%%%%%%%%%

\section{Introduction}

Classical novae (CNe) are binary systems \citep{1954PASP...66..230W,1964ApJ...139..457K} with a white dwarf (WD) accreting matter from a non-degenerate companion star (either main-sequence, sub-giant or red giant; see e.g.\ \citealp{2012ApJ...746...61D}). As material builds up on the surface of the WD the pressure and temperature increase until nuclear fusion occurs, leading to a thermonuclear runaway \citep{1972ApJ...176..169S}. This causes a rapid increase in luminosity, with the most luminous CN eruptions exceeding $M_V=-10$ (\citealp{2009ApJ...690.1148S}, Williams et al.\ in prep). By definition, novae with one observed eruption are classified as CNe; those with two or more observed eruptions are classified as recurrent novae (RNe). The shortest recurrence period observed to date is one year, in M31N 2008-12a (see e.g.\ \citealp{2016ApJ...833..149D}). For detailed reviews of the nova phenomenon see \citet{2008clno.book.....B} and \citet{2014ASPC..490.....W}. 

Novae have long been considered as potential Type Ia supernova (SN\,Ia) progenitor candidates (e.g.\ \citealp{1973ApJ...186.1007W}), with the latest models indicating that WDs in nova systems can indeed gain mass over a long series of eruption cycles and eventually produce a SN\,Ia \citep{2016ApJ...819..168H}. While it is widely accepted that SNe Ia are caused by thermonuclear explosions of carbon-oxygen WDs \citep{1960ApJ...132..565H,1984ApJ...286..644N,2000ARA&A..38..191H,2011Natur.480..344N}, the mechanisms via which the WD reaches the critical mass to explode is still unclear (see \citealp{2014ARA&A..52..107M} for a detailed review of SN\,Ia progenitor candidates). Although the production of lithium in nova eruptions has been predicted for some time (e.g.\ \citealp{1975A&A....42...55A,1978ApJ...222..600S,1996ApJ...465L..27H}), observational evidence has recently been found that novae may contribute the majority of the $^7$Li in the Galaxy \citep{2015Natur.518..381T,2016ApJ...818..191T,2015ApJ...808L..14I,2016MNRAS.463L.117M}.

While it is generally not possible to study individual extragalactic novae in as much detail as their Galactic counterparts, there are several advantages to observing extragalactic novae. The large uncertainties in distance that can be associated with Galactic novae are largely negated when studying extragalactic populations and a better representation is given of an entire galaxy's nova population. Additionally, it enables the studies of novae in different environments, for example the stellar populations of the nearby dwarf galaxies, the Large Magellanic Cloud (LMC) and Small Magellanic Cloud (SMC) are very different from the large spirals like M31 and our own Galaxy.

Many Local Group novae are discovered each year, yet to date detailed studies of individual nova eruptions in the low-metallicity environments typically found in dwarf irregular galaxies have been restricted to the nearby Magellanic Clouds (MCs). Dwarf galaxies of course have low nova rates, with the nova rates of the LMC and SMC calculated to be $2.4\pm0.8$\,yr$^{-1}$ and $0.9\pm0.4$\,yr$^{-1}$ respectively \citep{2016ApJS..222....9M}. This compares to rates of $65^{+16}_{-15}$\,yr$^{-1}$ in M31 \citep{2006MNRAS.369..257D}, $33^{+13}_{-8}$\,yr$^{-1}$ in M81 \citep{2004AJ....127..816N}, and even as high as $363_{-45}^{+33}$\,yr$^{-1}$ in M87 \citep{2016ApJS..227....1S}. To build a full picture of how the properties of novae depend on the properties of their host galaxy, it is important we study nova eruptions in these dwarf irregulars in as much detail as possible.

IC\,1613 is an irregular dwarf galaxy in the Local Group at a distance of approximately 730\,kpc \citep{2013ApJ...773..106S,2015MNRAS.452..910M}. Recent evidence suggests it has a metallicity of about one fifth of Solar, similar to that of the SMC \citep{2014ApJ...788...64G,2015MNRAS.449.1545B}, and its star formation rate has been constant over time \citep{2014ApJ...786...44S}. IC\,1613 differs from the MCs as it is essentially isolated, whereas the MCs are interacting with the Milky Way (see \citealp{2000glg..book.....V} for an overview).

A total of three nova candidates have previously been discovered in IC\,1613. The first was imaged at $B\simeq17.5$ on three plates taken on a single night by Walter Baade in 1954 November, having not been visible the night before and no further images were taken that season \citep{1971ApJ...166...13S}. The second candidate was detected on 1996 October 12, although the eruption time of this candidate is poorly constrained, with the last non-detection being two months prior \citep{2001A&A...367..759M}. For nine days following October 12, the candidate was seen to decline in brightness \citep{2001A&A...367..759M}. The third and most recent nova candidate was discovered in 1999 \citep{1999IAUC.7287....2K}, but this was actually a Mira variable \citep{2001A&A...378..449K}.

Nova IC\,1613 2015 (PNV J01044358+0203419) was discovered at $01^{\mathrm{h}}04^{\mathrm{m}}43^{\mathrm{s}}\!.58$~$+02^{\circ}03^{\prime}41^{\prime\prime}\!\!.9$ with an unfiltered magnitude of 17.5 on 2015 September 10.48\,UT, with nothing visible down to a limiting magnitude of about 18.0 on September 9 \citep{2015CBET.4186....1H}, by the Lick Observatory Supernova Search (see \citealp{2001ASPC..246..121F} for further details). After classification as an extragalactic nova \citep{2015ATel.8061....1W}, we conducted optical, near-IR, near-UV and X-ray observations of the eruption, which we present in this paper.

\section{Observations and Data Analysis}

\subsection{Ground-based photometry}

\begin{figure}
\includegraphics[width=\columnwidth]{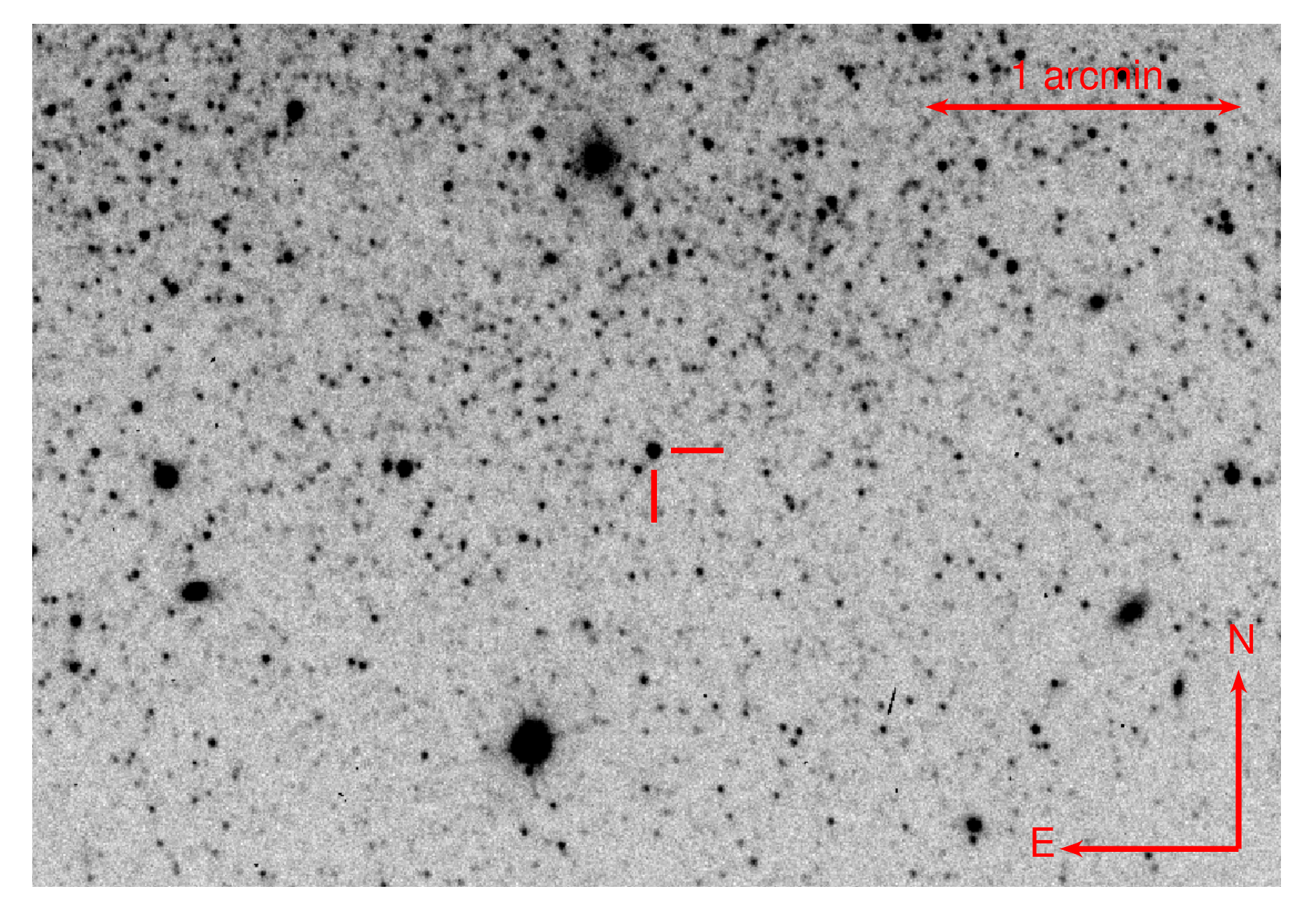}
\caption{Negative image of Nova IC\,1613 2015 in eruption taken though an {\it r}$^{\prime}$-band filter with IO:O on the LT on 2015 Oct 9.00\,UT. The position of the nova is indicated by the red lines near the centre of the image.\label{find}}
\end{figure}

Nova IC\,1613 2015 was initially followed with IO:O\footnote{\url{http://telescope.livjm.ac.uk/TelInst/Inst/IOO}}, the optical imager on the 2\,m Liverpool Telescope on La Palma, Canary Islands, Spain (LT; \citealp{2004SPIE.5489..679S}), using {\it B}, {\it V} and {\it i}$^{\prime}$ filters, with the first set of observations taken 1.61\,days after discovery on 2016 Sep 12.09\,UT. Once the nova nature of the object became clear, the filter set was expanded to {\it u$^{\prime}$}, {\it B}, {\it V}, {\it r}$^{\prime}$, {\it i}$^{\prime}$, and {\it z}$^{\prime}$. We also began monitoring the eruption in the near-IR using the fixed {\it H}-band filter on the IO:I imager on the LT \citep{2016JATIS...2a5002B}. In addition to the LT data, we also obtained some photometric observations through {\it B}, {\it V}, {\it r}$^{\prime}$, and {\it i}$^{\prime}$ filters using the Las Cumbres Observatory (LCO) 2\,m telescope at Siding Spring Observatory, New South Wales, Australia (formally the Faulkes Telescope South; FTS, \citealp{2013PASP..125.1031B}). An IO:O image of the nova in eruption is shown in Figure\,\ref{find}.

The {\it u$^{\prime}$}{\it BV}{\it r}$^{\prime}${\it i}$^{\prime}${\it z}$^{\prime}$ photometry was calculated using aperture photometry in GAIA\footnote{GAIA is a derivative of the Skycat catalogue and image display tool, developed as part of the VLT project at ESO. Skycat and GAIA are free software under the terms of the GNU copyright.} and calibrated against field stars from the Sloan Digital Sky Survey Data Release 9 \citep{2012ApJS..203...21A}. The {\it B} and {\it V} magnitudes of these calibration stars were calculated using the transformations in \citet{2006A&A...460..339J}. The {\it H}-band observations were calibrated against different stars from the 2MASS All Sky Catalog of point sources \citep{2003yCat.2246....0C}.

\subsection{Spectroscopy} \label{sec:spec}

The optical spectra were taken using the SPectrograph for the Rapid Acquisition of Transients (SPRAT), a low-resolution high-throughput spectrograph on the LT \citep{2014SPIE.9147E..8HP}. It has a 1$^{\prime\prime}\!\!.8$ slit width, giving a resolution of 18\,\AA. Our observations were all taken using the blue-optimised mode. The details of the spectra are summarised in Table\,\ref{tab:spec}.

\begin{table}
\caption{Summary of all spectroscopic observations of Nova IC\,1613 2015 with the SPRAT spectrograph on the LT.}              
\label{tab:spec}      
\begin{center}
\begin{tabular}{lcc}  
\hline                    
Date [UT]$^a$ & Days post-discovery & Exposure time [s]\\
\hline                                
2015 Sep 12.07 & 1.59 & 1800\\
2015 Sep 17.07 & 6.59 & 1800\\
2015 Sep 21.02 & 10.54 & 1800\\
2015 Sep 25.09 & 14.61 & 1800\\
2015 Oct 07.03 & 26.55 & 3600\\
2015 Nov 06.99 & 57.51 & 5400\\
\hline
\end{tabular}
\end{center}
$^a$~The date listed here refers to the mid-point of each observation.
\end{table}

Spectrophotometric standards were not observed at similar times as the IC\,1613 spectroscopy, but we observed the standard G191-B2B using the same SPRAT instrument set-up on 2015 Dec 17, 2015 Dec 30 and 2016 Jan 10. The flux calibration of each spectrum was performed using standard routines in IRAF\footnote{IRAF is distributed by the National Optical Astronomy Observatory, which is operated by the Association of Universities for Research in Astronomy (AURA) under a cooperative agreement with the National Science Foundation.} \citep{1986SPIE..627..733T}. The standard observations were calibrated against data from \citet{1990AJ.....99.1621O} obtained via ESO. Due to different observing conditions, and particularly seeing losses and atmospheric conditions (i.e.\ cloud), the absolute flux calibrations of each spectrum can vary significantly. Examining the standard star observations discussed above, we estimate this causes the typical flux calibration error (at 5000\,\AA) to be of order $15-20$\,\%. However, the relative calibration across any individual spectrum (i.e.\ between red and blue, after removing the systematic flux calibration offset) should be relatively good with the uncertainties $<10$\,\%.

\subsection{\textit{Swift} observations} \label{sec:swift}

The super-soft X-ray source (SSS) phase in novae is caused by nuclear burning of hydrogen on the surface of the WD. The SSS emission can be detected once the ejecta become optically thin to X-rays and the SSS `turn-off' is thought to represent the end of nuclear burning (see e.g.\ \citealp{1996ApJ...456..788K}).

We were granted six \swift\ \citep{2004ApJ...611.1005G} target of opportunity (ToO) observations (target ID 34085) to follow the UV and X-ray evolution of the nova. Additionally, we analysed data aimed at IC\,1613 itself (target ID 84201), which includes our object in the field of view. All \swift\ data are summarised in Table\,\ref{tab:swift}.

The \swift\ UV/optical telescope \citep[UVOT,][]{2005SSRv..120...95R} data were reduced using the HEASoft (v6.16) tool \texttt{uvotsource}. The UVOT magnitudes are based on aperture photometry of carefully selected source and background regions. The photometric calibration assumes the UVOT photometric (Vega) system \citep{2008MNRAS.383..627P} and have not been corrected for extinction. The central wavelengths of the utilised UVOT filters are: \textit{UVW1}: 2600\,\AA; \textit{UVM2}: 2250\,\AA; \textit{UVW2}: 1930\,\AA.

All \swift\ X-ray telescope \citep[XRT;][]{2005SSRv..120..165B} data were obtained in the photon counting (PC) mode. For extraction of the count rate upper limits we made use of the on-line interface\footnote{\url{http://www.swift.ac.uk/user\_objects}} of \citet{2009MNRAS.397.1177E}. This tool uses the Bayesian formalism of \citet{1991ApJ...374..344K} for low numbers of counts. As is recommended for SSSs, only grade zero events were extracted. To convert the counts to X-ray fluxes we assume a conservative (maximum) black-body temperature of 50\,eV and a Galactic foreground absorption of \nh = 3\hcm{20}. The absorption was derived from the HEASARC \nh tool based on the hydrogen maps of \citet{1990ARA&A..28..215D}.

We estimated the X-ray temperature based on the reference frame of the M31 SSS nova sample and the correlations subsequently found by \citet{2014A&A...563A...2H}. In M31, a t$_2$ of 13\,days (see Section\,\ref{s:phot}) would correspond to a SSS phase from about days 60--200, which in turn suggests a black-body $kT \sim 50$\,eV \citep[cf. figure 8 of][]{2014A&A...563A...2H}. Using the \texttt{pimms} software (v4.8c) we estimated an energy conversion factor (count rate divided by unabsorbed flux in the 0.2--1.0\,keV band) of 1.2\tpower{10}\,ct\,cm$^2$\,erg$^{-1}$ for the XRT (PC mode). We derived the corresponding X-ray luminosities in Table\,\ref{tab:swift} by assuming a distance to IC\,1613 of 730\,kpc.

\begin{table*}
\caption{\swift\ UVOT magnitude and X-ray upper limits.}
\label{tab:swift}
\begin{center}
\begin{tabular}{rrrrrrrrrr}
\hline
ObsID & Exp$^a$ & Date$^b$ & MJD$^b$ & $\Delta t^c$ & \multicolumn{3}{c}{UV$^d$ [mag]} & Rate &  L$_{0.2-1.0}^e$\\
& [ks] & [UT] & [d] & [d] & \textit{UVW1} & \textit{UVM2} & \textit{UVW2} & [\power{-3}\,ct\,s$^{-1}$] & [\power{37}\,erg\,s$^{-1}$]\\
\hline
00034085001 & 4.6 & 2015-10-20.54 & 57315.55 & 40.07 & $18.6\pm0.1$ & \ldots & \ldots & $<4.2$ & $<2.3$ \\
00034085002 & 4.2 & 2015-11-08.29 & 57334.29 & 58.81 & $19.6\pm0.1$ & \ldots & \ldots & $<2.6$ & $<1.5$ \\
00034085003 & 4.6 & 2015-11-29.22 & 57355.23 & 79.75 & $20.2\pm0.2$ & \ldots & \ldots & $<5.5$ & $<3.1$ \\
00034085004 & 4.1 & 2016-01-08.04 & 57395.05 & 119.57 & $20.9\pm0.3$ & \ldots & \ldots & $<2.1$ & $<1.2$ \\
00034085005 & 4.0 & 2016-02-09.09 & 57427.09 & 151.61 & $>20.9$ & \ldots & \ldots & $<3.2$ & $<1.8$ \\
00084201006 & 1.3 & 2016-02-17.81 & 57435.81 & 160.33 & $>19.5$ & $>19.8$ & $>20.0$ & $<8.7$ & $<4.8$ \\
00084201007 & 6.5 & 2016-05-26.43 & 57534.44 & 258.96 & $>20.1$ & $>20.8$ & $>21.1$ & $<1.8$ & $<1.0$ \\
00084201008 & 0.9 & 2016-05-28.90 & 57536.90 & 261.42 & $>19.3$ & $>19.6$ & $>19.8$ & $<9.5$ & $<5.3$ \\
00034085006 & 2.4 & 2016-08-07.07 & 57607.07 & 331.59 & $>21.0$ & \ldots & \ldots & $<4.9$ & $<2.7$\\
\hline
\end{tabular}
\end{center}
\begin{flushleft}
$^a$~Dead-time corrected XRT exposure time.\\
$^b$~Start date of the observation.\\
$^c$~Time in days after the eruption on 2015-09-10.48 UT (MJD 57275.48).\\
$^d$~Vega magnitudes for the $UVW1$, $UVM2$, and $UVW2$ filters with central wavelength: 2600\,\AA, 2250\,\AA, and 1930\,\AA, respectively.\\
$^e$~X-ray luminosity upper limits (unabsorbed, blackbody fit, 0.2--1.0\,keV) were estimated according to Sect.\,\ref{sec:swift}.
\end{flushleft}
\end{table*}

\subsection{Reddening}

IC\,1613 is subject to only a small amount of foreground reddening ($E_{B-V}=0.021$; \citealp{2011ApJ...737..103S}). However, estimating the reddening internal to IC\,1613 at the position of the nova is difficult, as this is highly variable throughout the galaxy \citep{2009A&A...502.1015G}. In a survey of IC\,1613 Cepheid variables, \citet{2006ApJ...642..216P} found an average total reddening of $E_{B-V}=0.090\pm0.019$ to the Cepheids, which we take as the extinction estimate for our absolute magnitude calculations.

\section{Results}

\subsection{Photometric evolution} \label{s:phot}

A light curve showing all the photometry taken by the LT, LCO 2\,m, and {\it Swift} is shown in Figure\,\ref{lc}. This photometry is also tabulated in Appendix\,\ref{append} and presented in the form of spectral energy distributions (SEDs) in Section\,\ref{sec:sed}. The light curve shows the nova was clearly discovered prior to peak.

\begin{figure}
\includegraphics[width=\columnwidth]{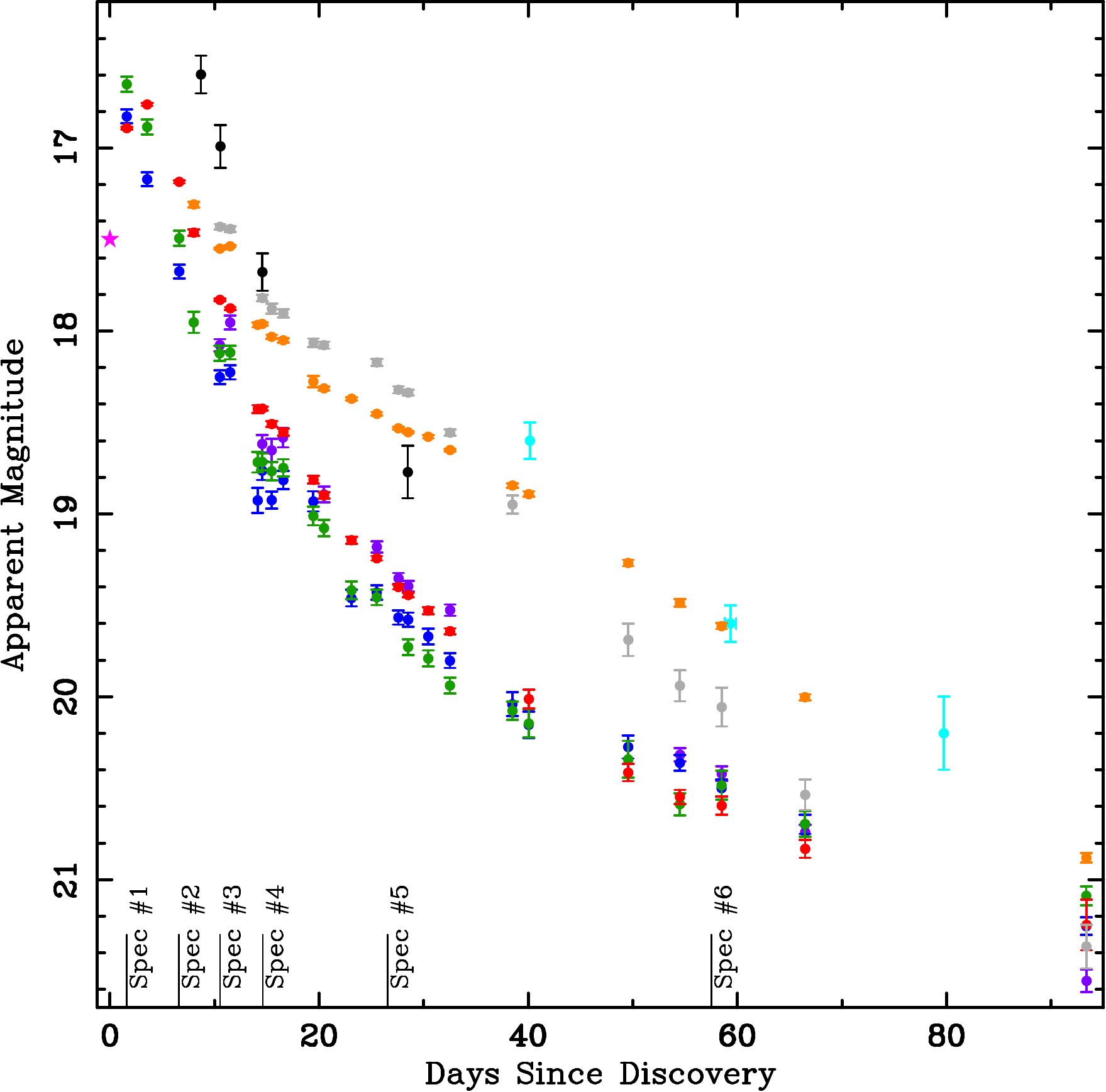}
\caption{Light curve of Nova IC\,1613 2015. The colours represent different filters: $UVW1$, cyan; {\it u$^{\prime}$}, purple; {\it B}, blue; {\it V}, green; {\it r}$^{\prime}$, orange; {\it i}$^{\prime}$, red; {\it z}$^{\prime}$, grey; {\it H}, black. The magenta star shows the unfiltered discovery magnitude. The points on the light curve that correspond to the dates the spectra were taken are also indicated.\label{lc}}
\end{figure}

The nova follows a relatively uniform decline, although the $r'$-band fades significantly slower than {\it B}, {\it V}, and $i'$ due to the increasingly strong influence of the H$\alpha$ emission line on the broadband photometry. Initially the nova also declines more slowly in the {\it z}$^{\prime}$-band than other filters, but by 40\,days post-discovery, the {\it z}$^{\prime}$-band declines more quickly than the other filters. The initial slow {\it z}$^{\prime}$ decline is unlikely to indicate a change in the overall nova SED, as the ($V - i'$) colour evolution remains relatively unaltered during this phase (the early {\it H}-band observations are also consistent) and therefore the slower {\it z}$^{\prime}$ decline is probably line driven. As we have no spectra extending beyond 8000\,\AA, the species that may be responsible for this is not certain, but we suggest it is most likely due to very strong O\,{\sc i} 8446\,\AA\ emission (caused by Ly$\beta$ fluorescence; see discussion in Section\,\ref{sec:spec_evo}).

Adopting a distance modulus $24.31\pm0.04$ (weighted average from \citealp{2013ApJ...773..106S,2015MNRAS.452..910M}) and correcting for reddening using $E_{B-V}=0.090\pm0.019$ \citep{2006ApJ...642..216P} and the extinction law from \citet[$R_V=3.1$]{1989ApJ...345..245C} gives an absolute magnitude for the eruption peak of $M_V=-7.93\pm0.08$, which is typical for a nova. The absolute magnitude at 15 days after peak is $M_V=-5.84^{+0.20}_{-0.10}$, we note that the (conservative) constraints on the time of peak (assuming 2016 Sep $12.09^{+1.95}_{-1.61}$\,UT) dominate the upper error bar. This is similar to that expected from the relationships of the absolute magnitudes of novae 15 (or 17) days after peak brightness in M49 ($M_{V,15}=-6.36\pm0.19$; \citealp{2003ApJ...599.1302F}) and M87 ($M_{F606W,17}=-6.06\pm0.23$; \citealp{2017arXiv170206988S}, although note the `wide {\it V}' F606W filter contains H$\alpha$). We measure the de-reddened day-15 colour, $(B-V)_{t=15}=-0.03^{+0.11}_{-0.14}$. In the SDSS filters we find $M_{r',15}=-6.50^{+0.17}_{-0.09}$, $(u'-r')_{t=15}=0.34^{+0.14}_{-0.09}$, $(r'-i')_{t=15}=-0.56\pm0.05$ and $(i'-z')_{t=15}=0.60^{+0.08}_{-0.05}$. The {\it H}-band coverage is much poorer than the other filters, but using extrapolation we estimate $M_{H,15}=-6.53^{+0.22}_{-0.20}$.

Taking the brightest data point as the peak magnitude of the nova, and using linear extrapolation between the data points, we estimate the $t_2$ of this nova to be $15\pm3$, $13\pm2$ and $15\pm3$\,days in {\it B}, {\it V} and $i'$ filters, respectively. We estimate the $t_3$ values to be $t_{3{\mathrm{(}}B{\mathrm{)}}}=32\pm3$, $t_{3{\mathrm{(}}V{\mathrm{)}}}=26\pm2$ and $t_{3{\mathrm{(}}i'{\mathrm{)}}}=32\pm3$\,days (the uncertainties here are largely due to the cadence around peak).

\subsection{Spectroscopic Evolution} \label{sec:spec_evo}

Novae have been observed spectroscopically for 150 years, since the first eruption of RN T\,Coronae Borealis in 1866 \citep{1866MNRAS..26..275H}. Nova spectra tend to fit into one of two groups, named after the dominant non-Balmer species in the spectra, Fe\,{\sc ii} and He/N classes \citep{1992AJ....104..725W}. These two types of spectra are suggested to form in different components of gas, with the spectral type observed for a given nova reflecting the dominant spectral component at that time \citep{2012AJ....144...98W}. The Fe\,{\sc ii} spectra have been suggested to originate in the circumbinary gas originating from the companion star, whereas the He/N type is suggested to be produced by the ejecta themselves \citep{2012AJ....144...98W}. Although there are some exceptions, Fe\,{\sc ii} novae tend to produce narrower emission lines than He/N novae (see e.g.\ \citealp{1992AJ....104..725W,2011ApJ...734...12S}). Line identification can be difficult in novae due to the broad lines and often differing line profiles. This is further complicated when using low-resolution spectra, which are often required to study faint extragalactic novae. However, the multiple epochs of spectra we have obtained allow us to identify some lines that may not have been possible with a single observation, and more importantly, better interpret the overall evolution. Line identification was also significantly aided by the extensive nova line list from \citet{2012AJ....144...98W} and multiplet tables from \citet{1945CoPri..20....1M}. The spectra are shown in Figures\,\ref{spec1}, \ref{spec2}, and \ref{spec6}. All spectra are shown in the frame of the observer, but when discussing the identification of spectroscopic features, rest-frame wavelengths are used. The average radial velocity of IC\,1613 is $-231.6$\,km\,s$^{-1}$, with a velocity dispersion of 10.8\,km\,s$^{-1}$ \citep{2014MNRAS.439.1015K}.

\subsubsection{Optically thick `fireball' stage}

\begin{figure*}
\includegraphics[width=2\columnwidth]{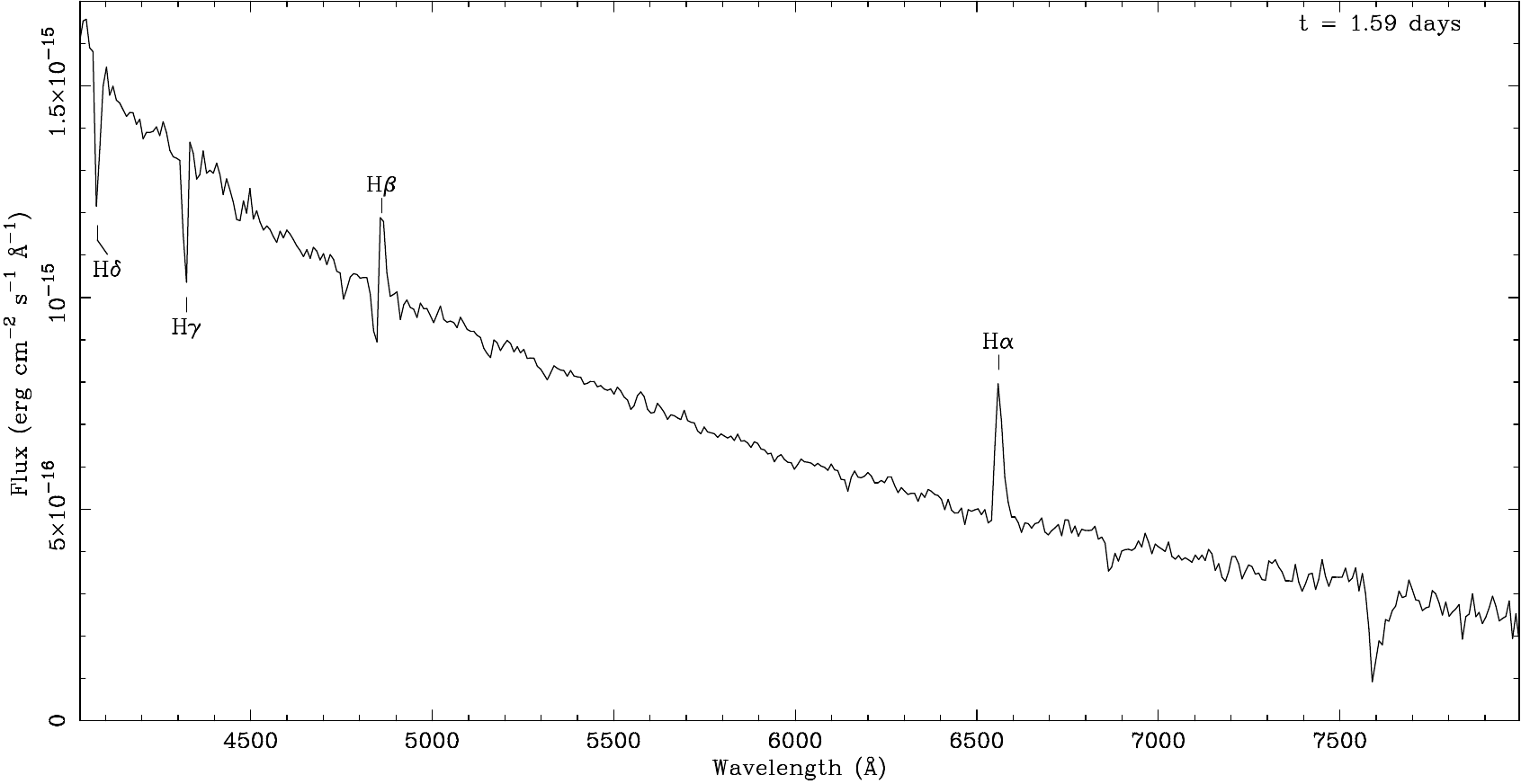}
\caption{The first spectrum of Nova IC\,1613 2015 taken 2015 Sep 12.07\,UT, 1.59\,days after discovery, and around peak optical brightness.\label{spec1}}
\end{figure*}

Our first spectrum was taken on 2015 Sep 12.07, 1.59\,days after discovery, and around peak brightness. The main features of this spectrum are the Balmer lines with clear P\,Cygni absorption profiles. H$\alpha$ is seen mainly in emission with a small, blue-shifted absorption component. H$\beta$ is seen with significant emission and absorption components, with H$\gamma$ and H$\delta$ mainly detected in absorption. This optically thick spectrum is shown in Figure\,\ref{spec1}. Fitting a Gaussian to the H$\gamma$ absorption profile and taking into account the radial velocity of IC\,1613 itself, the absorption minimum implies a velocity of $1200\pm200$\,km\,s$^{-1}$. Fe\,{\sc ii} 5169\,\AA\ is seen mainly in absorption, with features corresponding to the Fe\,{\sc ii} (42) triplet at 4924 and 5018\,\AA\ also tentatively detected. A few other weak absorption lines are also seen, e.g.\ one at 4465\,\AA\ from Mg\,{\sc ii}/Fe\,{\sc ii}.

\subsubsection{Early decline}

\begin{figure*}
\includegraphics[width=2\columnwidth]{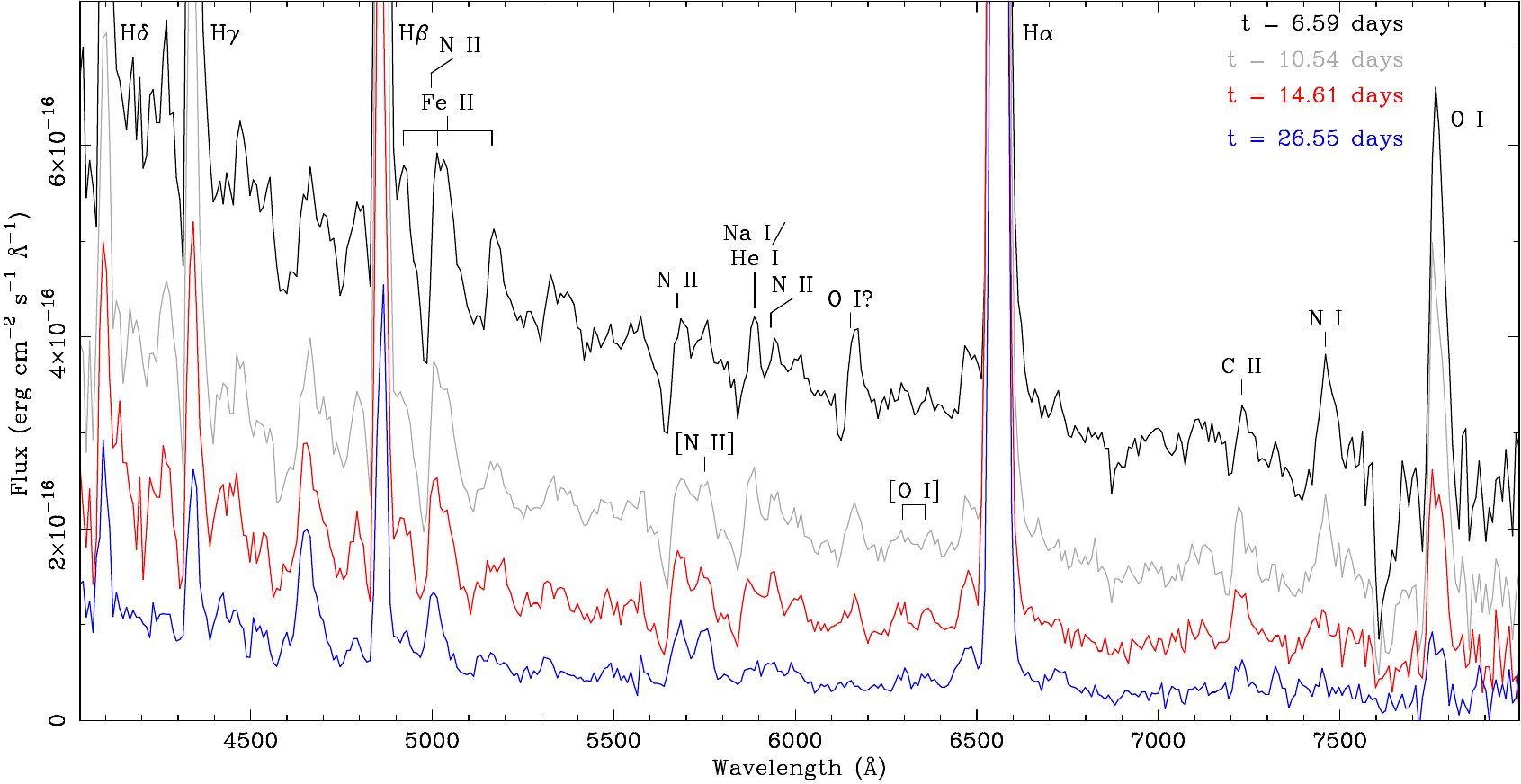}
\caption{The early decline spectra of Nova IC\,1613 2015. These were obtained 2015 Sep 17.07\,UT ($t=6.59$\,days; black line), Sep 21.02 ($t=10.54$\,days; grey line), Sep 25.09 ($t=14.61$\,days; red line) and Oct 7.03 ($t=26.55$\,days; blue line).\label{spec2}}
\end{figure*}

The second spectrum, taken 6.59\,days after discovery, shows a dramatic change from the $t=1.59$\,day spectrum and clearly shows the nova nature of the transient, indeed this is the spectrum we used to announce that the transient was a nova eruption in \citet{2015ATel.8061....1W}. This, along with the third, fourth and fifth spectra, taken at $t=10.54$, 14.61 and 26.55\,days, respectively, are shown in Figure\,\ref{spec2}. In the second spectrum, the nova now shows strong Balmer emission, although weak P\,Cygni absorption components are still present. The Fe\,{\sc ii} (42) triplet, only just detected in the first spectrum is now clearly seen in emission. In the early decline spectra of novae, triplet 42 is typically the strongest of the Fe\,{\sc ii} lines, although several other multiplets, located between H$\alpha$ and H$\beta$, are usually easily identifiable in regular Fe\,{\sc ii} novae (e.g.\ 48, 49, 55, and 74). The Fe\,{\sc ii} (48) multiplet is the only one of these that appears to be weakly detected in Nova IC\,1613 2015, with multiplets 49, 55, and 74 not detected.

In the $t=6.59$\,day spectrum, the N\,{\sc i} (3) triplet is seen strongly in emission. It also shows a relatively broad absorption component, as would be expected given the wavelengths of the lines that make up the triplet (7424, 7442, and 7468\,\AA), the velocities associated with the nova and the resolution of the spectrograph. The N\,{\sc ii} (3) multiplet around 5682\,\AA\ is seen with a very strong absorption component, with the N\,{\sc ii} (28) multiplet at 5938\,\AA\ also identified. The profile at the position of the Fe\,{\sc ii} 5018\,\AA\ component of triplet 42, clearly has a different morphology than the 5169\,\AA\ line, the former having a strong absorption component. We interpret this as indicating the presence of N\,{\sc ii} 5001\,\AA. This is consistent with the morphology of the N\,{\sc ii} (3) multiplet, which also shows a very strong absorption component. It also explains the evolution of the 5018\,\AA\ line profile between spectra two and five (see below). The Bowen blend (N\,{\sc iii}/C\,{\sc iii}/O\,{\sc ii}; this complex is discussed by Harvey et al.\ in prep), which is sometimes referred to as `4640 emission' (it is at $\sim$4640\,\AA) is detected as a broad emission line with an accompanying broad absorption profile. This complex is visible at the time of the emergence of the nebular lines in most novae, however is typically only visible in the early spectra if the nova is a member of the He/N spectroscopic class.

The strongest non-Balmer line visible in the $t=6.59$\,day spectrum is O\,{\sc i} 7774\,\AA, produced by O\,{\sc i} (1) triplet. There is a relatively strong emission line peaking at 6162\,\AA\ (again with an absorption profile). This is clearly not the Fe\,{\sc ii} (74) multiplet emission line at 6148\,\AA\ as the other lines are not present (notably the 6248\,\AA\ line, for example). We identify this as most likely being the O\,{\sc i} (10) triplet at 6157\,\AA. Alternatively it could be N\,{\sc ii} (which has lines at a similar wavelength), although that is perhaps less likely given it fades to be undetected by the fifth spectrum, which is very different from the other three N\,{\sc ii} lines (5001, 5682 and 5938\,\AA), which remain detected even in the final nebular spectrum, as discussed below. In the second spectrum, there is an emission line peaking at 6722\,\AA. An emission line at this wavelength has been noted since the early days of nova spectroscopy (e.g.\ DN\,Geminorum; \citealp{1940PLicO..14...27W}) and has been suggested as O\,{\sc i} 6726\,\AA\ (e.g.\ \citealp{1964PASP...76...22B}, \citealp{2005A&A...435.1031M}, \citealp{2013ATel.5378....1S}, \citealp{2014MNRAS.440.3402M}). An alternative explanation would be the N\,{\sc i} 6723\,\AA\ line.

The  Fe\,{\sc ii} 5018\,\AA/N\,{\sc ii} 5001\,\AA\ emission line appears to extend further redward than expected from other lines, which indicates the presence of another emission line. There is a He\,{\sc i} 5048\,\AA\ line which is a possible explanation, but there are no He\,{\sc i} lines at 6678 or 7065\,\AA, so a more likely explanation is N\,{\sc ii}. In the second and third spectra there is a weak feature at $\sim$7110\,\AA, which we identify as C\,{\sc i}. We also identify C\,{\sc ii} emission at 4267 and 7234\,\AA, which is visible from the day 6.59 to 26.55 spectra.

The spectroscopic evolution between day 6.59 and 26.55 shows the P\,Cygni profiles, that initially accompany many of the emission lines, weaken over time, as is usually seen in nova eruptions. The Fe\,{\sc ii} lines weaken along with the N\,{\sc i} and O\,{\sc i} lines (although O\,{\sc i} 7774\,\AA\ is still easily visible in the day\,26.55 spectrum). The N\,{\sc ii} and N\,{\sc iii} lines retain their strength through this evolution and by day\,26.55, apart from the Balmer and O\,{\sc i} 7774\,\AA\ lines, the spectrum is essentially dominated by ionised nitrogen lines (from the point of view of visible features, not the overall flux of the spectrum). The contrasting evolution of the Fe\,{\sc ii}/N\,{\sc ii} lines can be seen in the morphology of the blended line due to N\,{\sc ii} 5001\,\AA\ and Fe\,{\sc ii} 5018\,\AA, where the blue side of the blend become increasingly dominant as the nova evolves. 

\subsubsection{Nebular phase}

\begin{figure*}
\includegraphics[width=2\columnwidth]{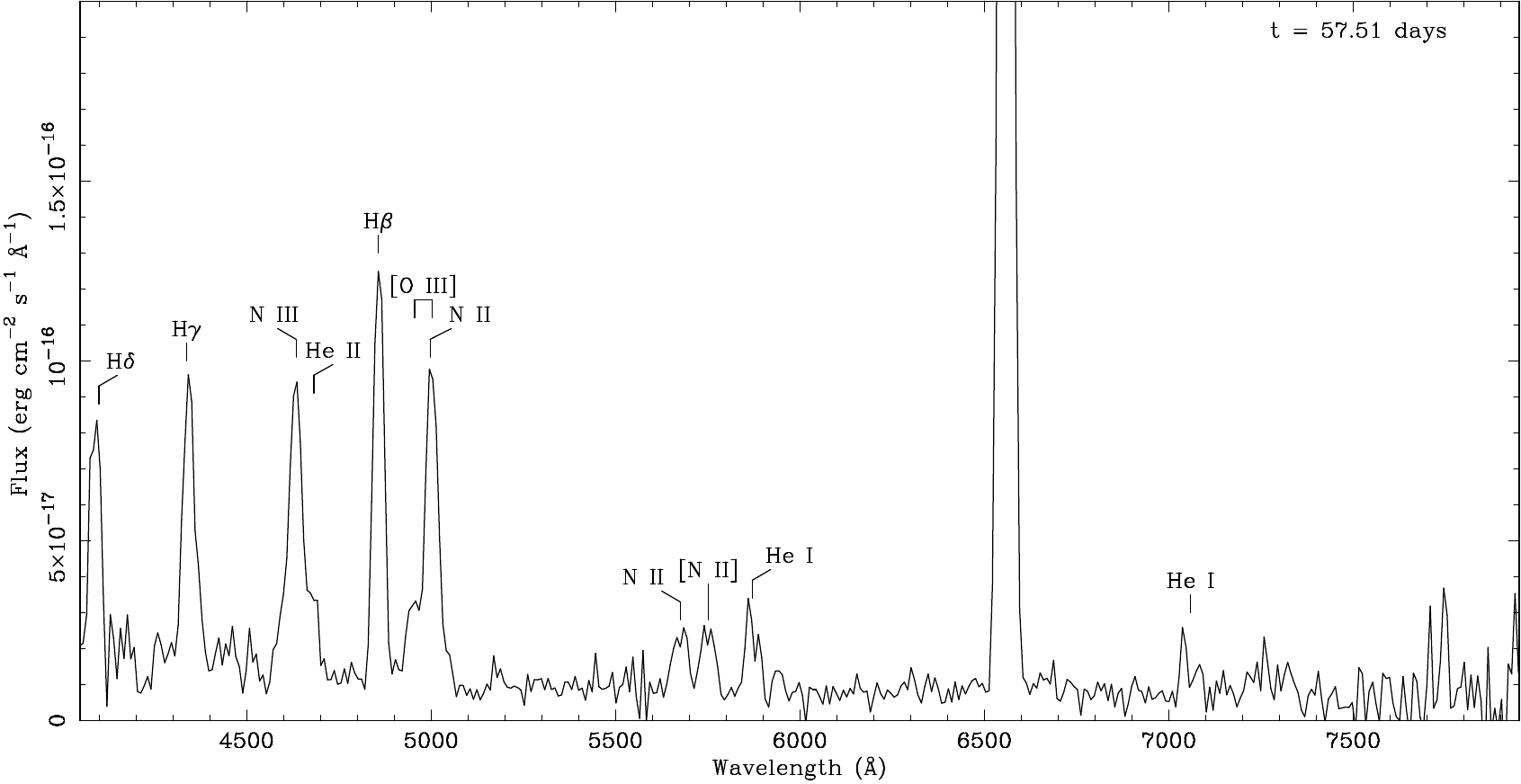}
\caption{The final spectrum of Nova IC\,1613 2015 taken 2015 Nov 6.99\,UT, 57.51\,days after discovery. This shows the nova in the nebular phase with strong [O\,{\sc iii}] emission.\label{spec6}}
\end{figure*}

Very few novae beyond the MCs have been observed spectroscopically in the nebular phase. As the evolution progresses, nebular lines also begin to emerge with [O\,{\sc i}] (6300 and 6364\,\AA) clearly detected by day 14.61 and possibly present even earlier. The [O\,{\sc ii}] 7320/7330\,\AA\ doublet is also likely present in the 26.55\,day spectrum.  Our sixth and final spectrum was taken 57.41\,days after discovery and about 4 magnitudes below peak. This shows further evolution into the nebular phase with [O\,{\sc iii}] (4959 and 5007\,\AA) now clearly visible. The 5007/4959\,\AA\ emission line ratio is higher than expected (should be $\sim3$; the lines are highly forbidden and the ratio is essentially fixed), indicating the line is still blended with N\,{\sc ii} 5001\,\AA. He\,{\sc ii} (4686\,\AA) can now be seen emerging from the Bowen complex, with the peak of the complex itself consistent with it being dominated by N\,{\sc iii}. We also find He\,{\sc i} emission (5876 and 7065\,\AA). This spectrum is shown in Figure\,\ref{spec6}. The emission line fluxes of prominent lines are shown in Table\,\ref{lflux}.

The assignment of [N\,{\sc ii}] (5755\,\AA) is correct in the later spectra (e.g.\ nebular [N\,{\sc ii}] would be expected when [O\,{\sc i}] is clearly detected in the $t=14.61$\,day spectrum), however there appears to be a line there even in the $t=6.59$\,day spectrum. This has been noted by other authors (e.g.\ \citealp{2003A&A...404..997I,2014AJ....147..107S}). There is also a non-forbidden N\,{\sc ii} doublet (9) at a very similar wavelength. This is caused by the same excited state as the N\,{\sc ii} (3) multiplet, but here the electrons transition to $2s^{2}2p3s$\,$^{1}$P$^{\circ}$, rather than $2s^{2}2p3s$\,$^{3}$P$^{\circ}$. It is possible that this at least partially contributes to this emission line.

\begin{table*}
\caption{The evolution of emission line fluxes.}              
\label{lflux}      
\begin{center}                                      
\begin{tabular}{lcccccc}  
\hline
Line identification & \multicolumn{6}{c}{Emission line flux [$\times10^{-15}$\,erg\,cm$^{-2}$\,s$^{-1}$]}\\               
(rest wavelength) & $t=1.59$\,days & $t = 6.59$\,days & $t = 10.54$\,days & $t = 14.61$\,days & $t = 26.55$\,days & $t = 57.51$\,days\\ 
\hline
H$\delta$ (4102\,\AA) &\ldots &$18.7\pm3.3$ &$9.5\pm1.2$ &$7.0\pm1.9$ &$4.3\pm1.1$ &$3.9\pm0.6$\\
H$\gamma$ (4341\,\AA) &\ldots &$28.5\pm5.2$ &$16.6\pm3.0$ &$9.0\pm1.6$ &$4.9\pm0.7$ &$4.4\pm0.3$\\
H$\beta$ (4861\,\AA) &$3.2\pm1.6$ &$68.0\pm4.6$ &$43.9\pm4.0$ &$24.6\pm2.9$ &$12.1\pm1.2$ &$5.4\pm0.3$\\
Fe\,{\sc ii} (5169\,\AA) &\ldots &$3.8\pm0.8$ &$2.7\pm0.4$ &$2.7\pm0.6$ &\ldots &\ldots\\ 
N\,{\sc ii} (5682\,\AA) &\ldots &$1.5\pm0.5$ &$2.8\pm0.7$ &$3.2\pm0.5$ &$3.1\pm0.3$ &$1.0\pm0.2$\\\relax   
[N\,{\sc ii}] (5755\,\AA) &\ldots &$1.2\pm0.2$ &$2.6\pm0.3$ &$2.5\pm0.4$ &$3.5\pm0.3$ &$1.0\pm0.2$\\      
N\,{\sc ii} (5939\,\AA) &\ldots &$1.1\pm0.2$ &$2.4\pm0.6$ &$2.2\pm0.3$ &\ldots &\ldots\\ 
O\,{\sc i} (6157\,\AA) &\ldots &$2.6\pm0.3$ &$1.8\pm0.3$ &$1.1\pm0.3$ &\ldots &\ldots\\\relax 
[O\,{\sc i}] (6300\,\AA) &\ldots &\ldots &\ldots &$1.0\pm0.3$ &$0.6\pm0.2$ &\ldots\\\relax  
[O\,{\sc i}] (6364\,\AA) &\ldots &\ldots &\ldots &$0.9\pm0.2$ &\ldots &\ldots\\
H$\alpha$ (6563\,\AA) &$7.3\pm1.1$ &$146.4\pm5.1$ &$144.0\pm5.2$ &$122.0\pm3.7$ &$90.5\pm7.4$ &$29.1\pm2.1$\\
C\,{\sc ii} (7235\,\AA) &\ldots &$1.4\pm0.7$ &$2.1\pm0.7$ &$2.7\pm0.4$ &$1.2\pm0.3$ &\ldots\\
N\,{\sc i} (7452\,\AA) &\ldots &$5.4\pm0.7$ &$3.2\pm0.4$ &$1.6\pm0.5$ &\ldots &\ldots\\
O\,{\sc i} (7774\,\AA) &\ldots &$18.1\pm2.2$ &$16.6\pm2.7$ &$8.5\pm1.7$ &$2.8\pm1.2$ &\ldots\\
\hline
\end{tabular}
\end{center}
\begin{flushleft}
The emission line fluxes are dependent on the assumed continuum level and if a P\,Cygni absorption component is present, only the emission component of the feature is measured. The fluxes are dereddened for foreground Galactic extinction, assuming $E_{B-V}=0.021$ and $R_V=3.1$. The errors do not take into account uncertainties in the flux calibration discussed in Section\,\ref{sec:spec}.
\end{flushleft}
\end{table*}

\subsubsection{Balmer evolution} \label{sec:balmer}

By fitting of a Gaussian profile to the emission component of the H$\alpha$ line in the second ($t = 6.59$\,days) spectrum, we measure the FWHM to be $1580\pm70$\,km\,s$^{-1}$ after correcting for the spectral resolution. The line then apparently broadens to a FWHM of $1750\pm50$\,km\,s$^{-1}$ in the third ($t=10.54$\,day) spectrum and thereafter remains consistent. We measure it at $1760\pm90$, $1750\pm120$ and $1720\pm190$\,km\,s$^{-1}$ in the $t = 14.61$, $t = 26.55$ and $t = 57.51$\,day spectra, respectively. The most likely explanation for the early change in the FWHM is that in the $t = 6.59$\,day spectrum the H$\alpha$ emission line is significantly influenced by a P\,Cygni absorption component, which has the effect of narrowing the apparent emission line.

\begin{figure*}
\includegraphics[width=\columnwidth]{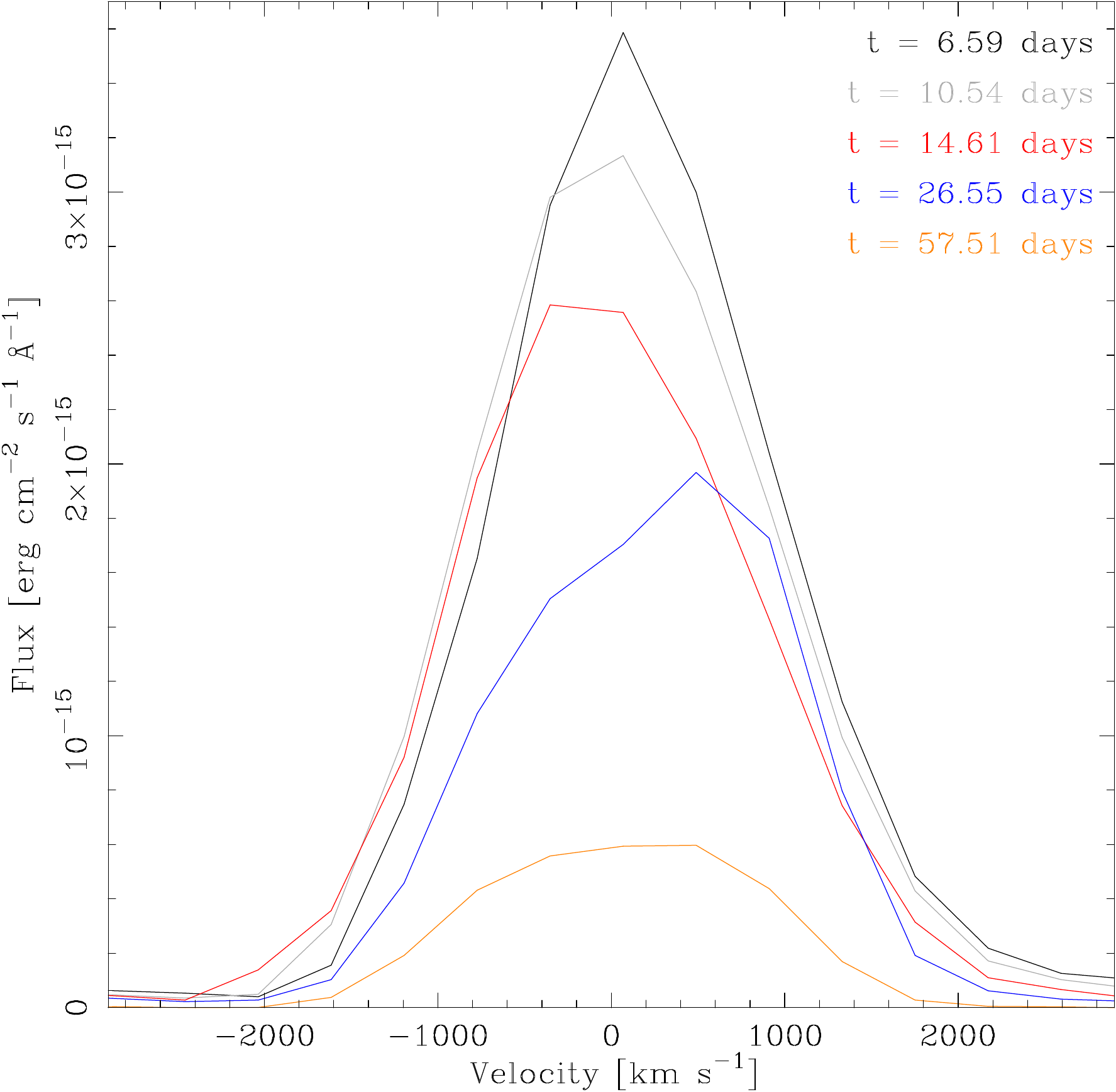}\hfill
\includegraphics[width=\columnwidth]{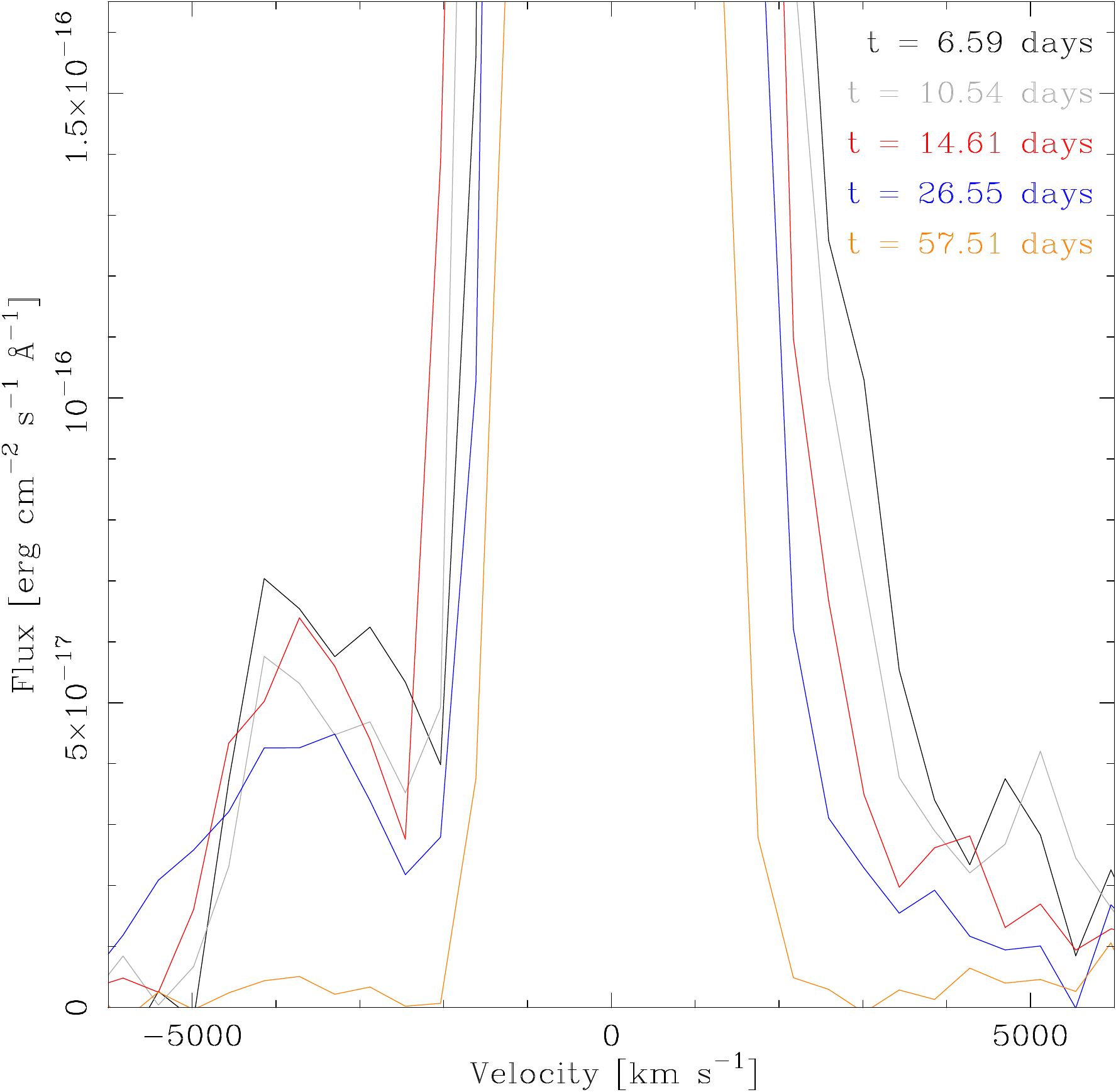}
\caption{Evolution of the H$\alpha$ line from $t=6.59$ to $t=57.51$\,days. The left panel shows the evolution of the shape and absolute flux of the line. The right panel shows the evolution of the fainter emission either side of the main H$\alpha$ profile (see discussion in Section\,\ref{sec:balmer}). The velocities have been corrected for the radial velocity of IC\,1613 itself.\label{halpha}}
\end{figure*}

The peaks of the Balmer emission are not shown in Figures\,\ref{spec2} and \ref{spec6} to allow the reader to view the fainter lines in greater detail. The evolution of the H$\alpha$ line is show in Figure\,\ref{halpha}. The left panel of Figure\,\ref{halpha} shows the overall profile is relatively symmetrical, with a Gaussian profile generally fitting the central profile well. The only stage when a Gaussian does not appear a good fit is the $t=26.65$\,day spectrum, when the profile seems asymmetric, being stronger at the red side of the profile. Close inspection of the H$\alpha$ profile in the right panel of Figure\,\ref{halpha} shows there is emission peaking at around $-4000$\,km\,s$^{-1}$ ($\sim$6480\,\AA). Comparing it to the red side of the H$\alpha$ line shows it is too blue to be explained as part of a simple Gaussian with a P\,Cygni profile superimposed on the emission component, and could be due to emission from N\,{\sc i} or N\,{\sc ii}. An alternative explanation could be a separate higher velocity component to the H$\alpha$ line as there may be excess flux on the red side of the profile as well, although this $\sim$6480\,\AA\ emission appears to persist longer, visible in all but the final nebular spectrum.

\begin{figure}
\includegraphics[width=\columnwidth]{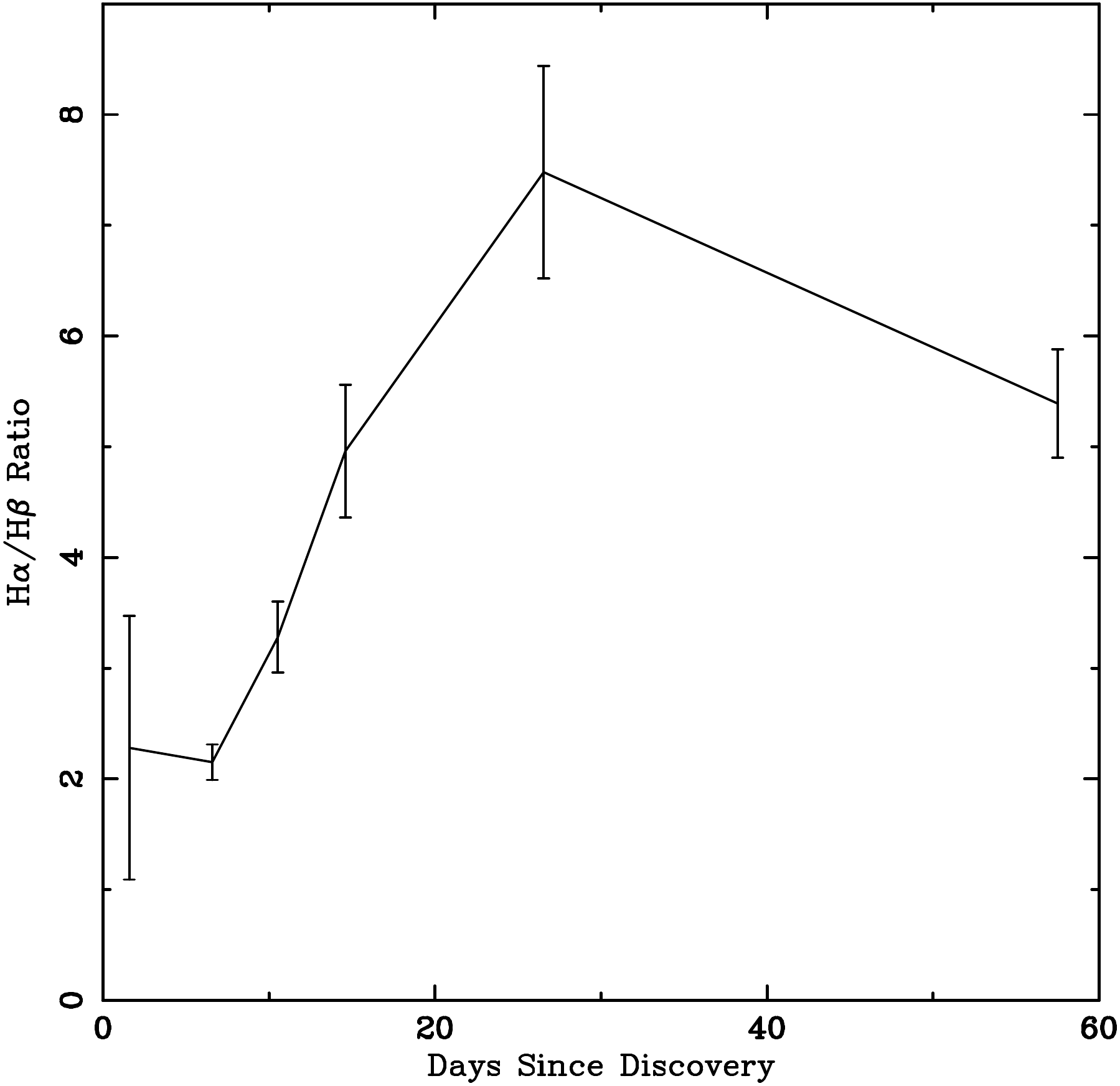}
\caption{The evolution of the H$\alpha$/H$\beta$ intensity ratio between the $t=6.59$ and $t=57.51$\,day spectra. The ratios are corrected for foreground Galactic reddening ($E_{B-V}=0.021$).\label{dec}}
\end{figure}

In Figure\,\ref{dec} we show the evolution of the H$\alpha$/H$\beta$ ratio between $t=6.59$ and $t=57.51$\,days, corrected for Galactic reddening ($E_{B-V}=0.021$). The figure shows the ratio initially increases, peaking at $7.5\pm1.0$ on $t = 26.55$\,days, before declining in the final nebular phase spectrum. During this period the H$\gamma$/H$\beta$ ratio does not change dramatically. The evolution in the H$\alpha$/H$\beta$ ratio can clearly not be caused by dust as such a dramatic change would be seen as a dip in the optical light curve. This H$\alpha$/H$\beta$ ratio evolution is common in novae and has been discussed by a number of authors (e.g.\ \citealp{1961PASJ...13..335K,1963ApJ...137..834M,1978ApJ...219..589F,1979ApJ...232..382F,1992A&A...263...87A,2003A&A...404..997I}). The changing Balmer decrement is caused by self-absorption. If Ly$\alpha$ and H$\alpha$ have high optical depth, high H$\alpha$/H$\beta$ ratios such as those observed here can be produced \citep{1975MNRAS.171..395N}. The calculations made by \citet{1975MNRAS.171..395N} also indicate the H$\gamma$/H$\beta$ ratio does not necessarily change dramatically during this H$\alpha$/H$\beta$ evolution, although this is dependent on the Ly$\alpha$ optical depth. This Balmer line ratio behaviour appears to well replicate  that observed in Nova IC\,1613 2015. In the case of novae in eruption, Case B recombination is not valid (as discussed above), therefore the Balmer decrement cannot, and should not, be used to estimate reddening.

As suggested by \citet{1947PASP...59..196B}, the close proximity of O\,{\sc i} 1025.76\,\AA\ to Ly$\beta$ (1025.72\,\AA) can lead to excitation of the O\,{\sc i} ground state. This then produces strong emission at the 1.1287\,$\mu$m and 8446\,\AA\ wavelengths as the electrons fall back to the O\,{\sc i} ground state (see also \citealp{1995ApJ...439..346K}). This effect and its relationship to the Balmer decrement has also been discussed in the context of Seyfert galaxies \citep{1974ApJ...191..309S}. Such Ly$\beta$ florescence can only occur under conditions of optically thick hydrogen. The H$\alpha$/H$\beta$ ratio and the O\,{\sc i} 8446\,\AA\ intensity are closely linked \citep{1979ApJ...229..274F,1979ApJ...232..382F}, with the H$\alpha$/H$\beta$ and (O\,{\sc i}~8446\,\AA)/H$\beta$ line ratios often peaking at a similar point of the eruption (see e.g.\ \citealp{1986ApJS...60..375F}). We note that between the 26.55 and 57.51\,day spectra the H$\alpha$/H$\beta$ ratio drops. Between the 26.55 and 57.51\,days spectra, it can also be seen from Figure\,\ref{lc} that the {\it z}$^{\prime}$-band fades more rapidly than any other waveband, indicating Ly$\beta$ florescence (or specifically the 8446\,\AA\ line; as indicated by the drop in the H$\alpha$/H$\beta$ ratio) may be the cause of the initially slower {\it z}$^{\prime}$-band decline mentioned in Section\,\ref{s:phot}.

\subsection{Spectral energy distributions} \label{sec:sed}

SEDs can be derived from multiband photometry or spectra, both of which have drawbacks. Spectra are more time-expensive and are more prone to (variable and colour-dependent) calibration issues. However, they allow prominent spectral features not associated with the underlying SED to be removed before fitting, which broadband photometric observations do not. This is particularly important in novae, where during an eruption, the broadband photometry becomes increasingly influenced by line emission and can even be dominated by it at late times (e.g.\ [O\,{\sc iii}] and H$\alpha$).

Fitting a power-law to the first spectrum (excluding prominent emission and absorption features) indicates at 1.59\,days post-discovery, $f_{\lambda}\propto\lambda^{-2.42\pm0.08}$ at optical wavelengths. This is near that expected from optically thick free-free emission ($f_{\lambda}\propto\lambda^{-8/3}$; \citealp{1975MNRAS.170...41W}). The relatively short wavelength coverage however is also consistent with a black-body. Fitting a black-body function to the first spectrum yields a photospheric temperature of $11600\pm500$\,K. This is close to the expected effective temperature of a nova at peak ($\sim8000$\,K; see e.g.\ \citealp{1986ApJ...310..222P,1995A&A...294..195B,2005MNRAS.360.1483E}). Five days later the wavelength dependence of the optical continuum had changed dramatically to $f_{\lambda}\propto\lambda^{-1.48\pm0.08}$, even shallower than that expected from optically thin free-free ($f_{\lambda}\propto\lambda^{-1.9}$). The measured wavelength dependence of the SED increases for the third (10.54\,day) spectrum, with $f_{\lambda}\propto\lambda^{-1.58\pm0.09}$, although note these are consistent within the errors. The other three spectra give $\propto\lambda^{-1.68\pm0.11}$, $\appropto\lambda^{-2.11\pm0.12}$ and $\propto\lambda^{-1.21\pm0.18}$ on days 14.61, 26.55 and 57.51, respectively. Note that these fits only include the known foreground reddening ($E_{B-V}=0.021$), therefore the intrinsic slope of the SEDs of the nova eruption could be bluer.

\begin{figure}
\includegraphics[width=\columnwidth]{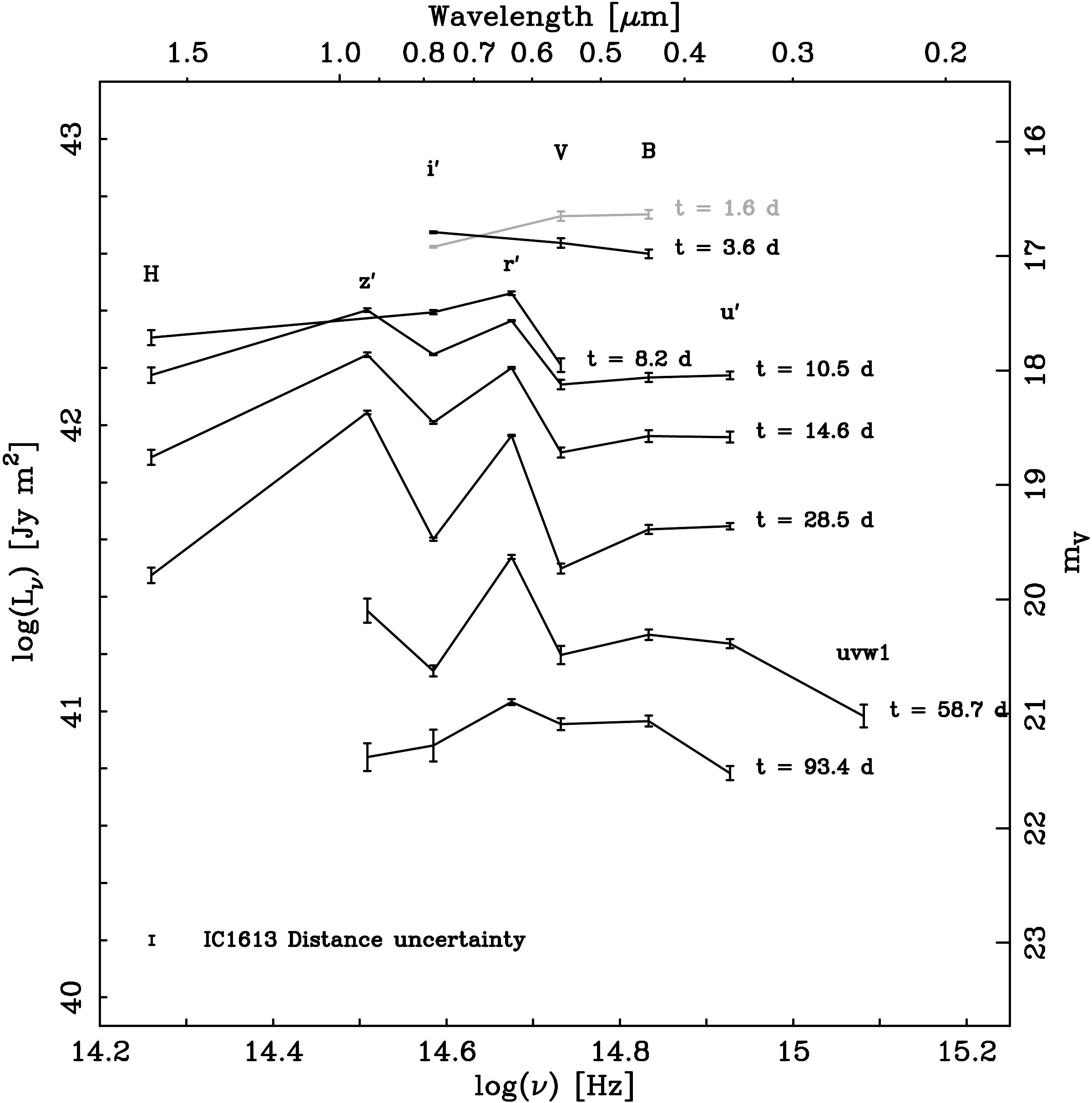}
\caption{SEDs of Nova IC\,1613 2015 from peak to 93.4\,days post-discovery (around $m_V=4.4$ below peak). The extreme effect of the H$\alpha$ emission on the {\it r}$^{\prime}$-band photometry can easily be seen. The size of the systematic uncertainty from the distance of IC\,1613 is indicated near the bottom of the plot.\label{sed}}
\end{figure}

The SEDs from the multi-band photometry are shown in Figure\,\ref{sed}. The photometry taken at similar epochs as the spectra discussed above are broadly consistent with the power-laws derived from the spectra themselves, keeping in mind that as the nova fades, the broadband photometry becomes increasingly influenced by strong emission lines (e.g.\ Balmer and O\,{\sc i}, and later [O\,{\sc iii}]). The photometry undoubtedly confirms that the dramatic change in the slope of the optical continuum emission between the spectra on day 1.59 and day 6.59 is real. Indeed a large change occurs between day 1.6 (i.e.\ the time of the first spectrum) and day 3.6.

\subsection{X-rays}
We do not detect X-ray emission from the nova between days 40--330 after eruption. The resulting luminosity upper limits, listed in Table\,\ref{tab:swift} for each individual \swift\ observation, were typically below 5\ergs{37}. This allows us to rule out a bright SSS under the conservative assumption of a 50\,eV black-body spectrum \citep[the fastest novae are considerably hotter, see e.g.][]{2011ApJ...727..124O,2014A&A...563L...8H,2015MNRAS.454.3108P}. Before day 40, the nova was still bright in UV and no X-rays would have been emitted \citep[cf.][]{2015A&A...580A..46H}. An observing gap between days 160--260 (Table\,\ref{tab:swift}) is likely too short to hide a SSS phase: based on our experience with the M31 population, a nova with a SSS turn-on time of more than 160 days would be expected to be visible in X-rays for longer than 260 days \citep{2014A&A...563A...2H}.

Combining all observations in Table\,\ref{tab:swift} (32.6\,ks) we derive an upper limit of 4.5\cts{-4}, corresponding to a luminosity of 2.4\ergs{35}. This is an order of magnitude lower than the observed luminosities of faint novae in M31 \citep{2010A&A...523A..89H,2011A&A...533A..52H,2014A&A...563A...2H}. Since fainter novae are typically visible for longer \citep{2014A&A...563A...2H} we can rule out a low-luminosity SSS counterpart for the observed time range. Note that this upper limit would only be valid if such a low luminosity SSS was emitting over the whole time frame of the {\it Swift} observations.

\subsection{The Progenitor Search}

The position of Nova IC\,1613 2015 is not covered by {\it Hubble Space Telescope} data (which is ideal for such progenitor searches due to its high resolving power and large wavelength coverage), however at the distance of IC\,1613, the most luminous quiescent systems will still be detectable in deep ground-based data. As we noted in \citet{2015ATel.8061....1W}, the nova appears very close to a {\it V} = 22.06, {\it I} = 21.53\,magnitude source recorded at $01^{\mathrm{h}}04^{\mathrm{m}}43^{\mathrm{s}}\!.56$~$+02^{\circ}03^{\prime}42^{\prime\prime}\!\!.0$ (J2000) in \citet{2001AcA....51..221U}.

\begin{figure*}
\includegraphics[width=\columnwidth]{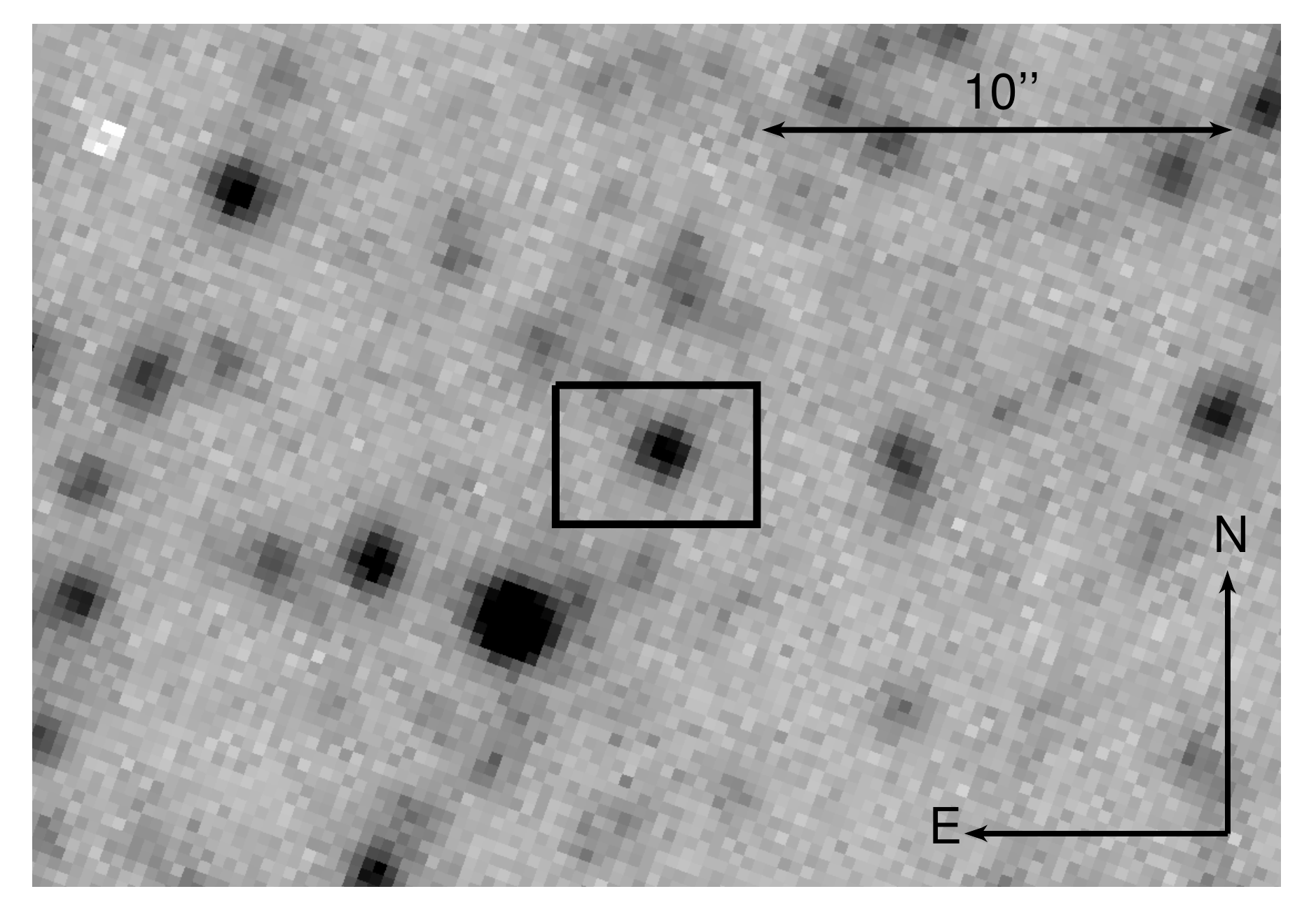}\hfill
\includegraphics[width=\columnwidth]{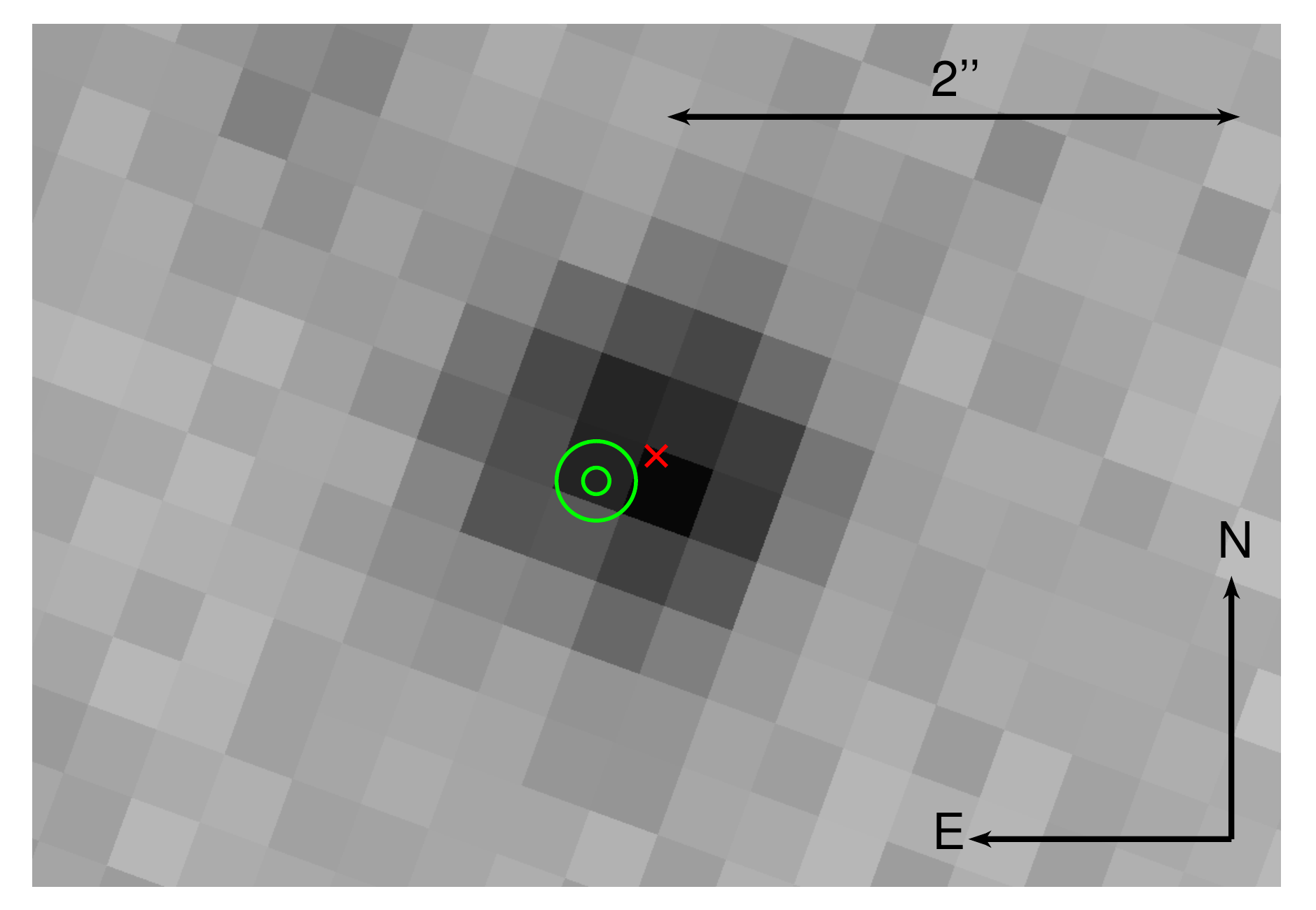}
\caption{The position of the nova in pre-eruption data compared to the nearby resolved source. Left: The Nova IC\,1613 2015 nova field imaged with FORS2 on the VLT using an R\_SPECIAL filter on 2012 Aug 20. The black box indicates the zoomed in region shown in the right panel. Right: The same data as the left panel, with the 1$\sigma$ and 3$\sigma$ errors on the position of the nova (calculated from $r'$-band LT eruption data) indicated by green circles and the position of the nearby resolved source indicated by a red `$\times$'. 
\label{prog1}}
\end{figure*}

The field was observed with the Very Large Telescope (VLT) using the FOcal Reducer/low dispersion Spectrograph 2 (FORS2; \citealp{1998Msngr..94....1A}) on 2012 Aug 20 under proposal 090.D-0009(A) and using the R\_SPECIAL filter (effective wavelength 6550\,\AA). Using the method described in detail by \citet{2009ApJ...705.1056B}, \citet{2014A&A...563L...9D} and \citet{2014ApJS..213...10W}, we used reference stars in {\it r}$^{\prime}$-band LT eruption images to precisely determine the position of the nova in the archival data. This is shown in Figure\,\ref{prog1}. We also independently (using different reference stars) calculated the position of the nova in archival SDSS {\it g}-band OmegaCAM \citep{2002Msngr.110...15K,2011Msngr.146....8K} data taken at the 2.6\,m VLT Survey Telescope \citep{1998Msngr..93...30A} on 2014 Dec 17, using {\it V}-band LT data.

In the first archival image where the position was derived from the {\it r}$^{\prime}$-band LT eruption observations, the position of the nova is calculated to be $0^{\prime\prime}\!\!.09\pm0^{\prime\prime}\!\!.05$\,south and $0^{\prime\prime}\!\!.21\pm0^{\prime\prime}\!\!.05$\,east of the progenitor candidate. Using the errors on the positional transformation and the centroid on the nova/stellar source, implies an association between the two sources can be ruled out at the 4.1$\sigma$ level. In the second image we calculate the position of the nova to be $0^{\prime\prime}\!\!.05\pm0^{\prime\prime}\!\!.05$\,north and $0^{\prime\prime}\!\!.25\pm0^{\prime\prime}\!\!.05$\,east. From this it appears the progenitor candidate may have a small, but real offset (eastward) from the nova.

As a check for a systematic offset across the transformed field, we apply the positional transformation to 10 stars in close proximity to the nova (within $\sim$1$^{\prime}$), that were not used in the calculation of the positional transformation itself. There is no evidence for a systematic offset in these sources, with the average x/y offsets less than the standard deviation in the offsets. The standard deviation of these offsets however does indicate that it is possible the errors on the transformation are slightly underestimated. We therefore apply the average x/y offsets (in the R\_SPECIAL image) to the position of the nova. Using the standard deviation as the error indicates an association is still ruled out, but with a reduced 3$\sigma$ confidence. It is also worth noting that the transformed position of the source to the south-east of the nova (the brightest star seen in the left panel of Figure\,\ref{prog1}) is consistent within 1.1$\sigma$ (using the errors of the transformation itself) of that of the centroided position from the VLT image. If there were a systematic offset affecting the nova transformation, one would expect it to also be present in this very nearby ($\sim$5$^{\prime\prime}$ separation) source.

We therefore conclude that, despite the close proximity of the progenitor candidate, it is most likely simply a chance alignment. However, this should be confirmed by late-time spectroscopy. Novae retain strong Balmer emission for a significant time after eruption, but over time the optical spectrum becomes increasingly dominated by [O\,{\sc iii}] emission lines. If this progenitor candidate is indeed the luminous accretion disk of the pre-eruption nova, a quiescent spectrum may be expected to reveal narrow Balmer and He\,{\sc ii} emission. The lack of (broader) [O\,{\sc iii}] lines would confirm we are not observing an extended tail of the nova eruption.

\section{Discussion}

A comparison of Nova IC\,1613 2015 with other IC\,1613 novae is not possible due to the lack of data for the 1954 and 1996 candidates. We can however compare it to other extragalactic novae residing in Local Group galaxies.

At $t_{2(V)}=13\pm2$\,days, Nova IC\,1613 2015 can be considered a fast-fading nova \citep{1957gano.book.....G}. Comparing it to the cumulative $t_2$ distribution plot of M31 novae from \citet{2016ApJ...817..143W} would place it in the fastest 20\% of novae. However a better comparison may be the LMC, and comparing the $t_2$ value to those in Table\,2 from \citet{2013AJ....145..117S} reveals that in this (albeit small) sample, novae significantly slower than Nova IC\,1613 2015 are relatively rare. The low nova rate of the SMC, perhaps the best comparison to IC\,1613, makes a comparison to the overall SMC population difficult. There are several SMC novae with good light curves (see e.g.\ \citealp{1954PNAS...40..365H,1998A&A...335L..93D,2016ApJS..222....9M}) which display a broad range of decline times, and Nova IC 2016 2015 would certainly not seem out of place amongst these. Indeed, there have been some novae that evolved much more slowly than Nova IC\,1613 2015 (e.g.\ Nova SMC 1994, \citealp{1998A&A...335L..93D}; Nova SMC 2001, \citealp{2004IBVS.5582....1L,2016ApJS..222....9M}).

The early decline spectra of Nova IC\,1613 2015 are not typical of novae. In M31, around 80\% of all novae belong to the Fe\,{\sc ii} class \citep{2011ApJ...734...12S}. From the smaller sample size of LMC and M33 novae, a lower proportion (perhaps around 50\%) appear to be Fe\,{\sc ii} novae in these galaxies \citep{2012ApJ...752..156S,2013AJ....145..117S}. The hybrid spectroscopic class of novae can either evolve from one type to another or show both types simultaneously. Nova IC\,1613 2015 shows both Fe\,{\sc ii} lines and N\,{\sc ii} in the early decline spectra, classifying it as a hybrid nova. It is worth noting that it is not unreasonable to expect hybrid novae that transition from one type to another to go through a phase which simultaneously shows both types, even if it is only short lived. The early decline spectral evolution shows many similarities to the hybrid Nova LMC 1988 No.\,2 (see \citealp{1989MNRAS.241..827S} and \citealp{1991ApJ...376..721W}), although Nova IC\,1613 2015 fades significantly more slowly, with Nova LMC 1988 No.\,2 having a $t_2$ of around 5\,days \citep{1989MNRAS.241..827S}. The evolution of Nova IC\,1613 2015 also appears similar to the Galactic nova V5114\,Sagittarii, which had a similar $t_2$ ($\sim 11$\,days) and showed somewhat similar spectroscopic evolution and associated velocities (H$\alpha$ FWHM $\sim 2000$\,km\,s$^{-1}$; \citealp{2006A&A...459..875E}), although in this case the spectrum shortly after peak appears closer to a typical Fe\,{\sc ii} nova \citep{2006A&A...459..875E} than Nova IC\,1613 2015.

As the best Local Group galaxy with observed novae to compare IC\,1613 to is the SMC, it is worth reviewing the spectroscopic information on SMC novae. Nova SMC 1951 was observed spectroscopically several times and clearly shows the Bowen blend emission complex \citep{1954PNAS...40..365H}. However most novae show this at later times, so an unambiguous spectroscopic type cannot be assigned. Nova SMC 1952 was observed two days after peak, and likely showed He\,{\sc i} and He\,{\sc ii} emission \citep{1954AJ.....59R.193S}, classifying it as an He/N nova. Nova SMC 2001 was an Fe\,{\sc ii} nova \citep{2005A&A...435.1031M}. Nova SMC 2005 shows broad Fe\,{\sc ii} lines \citep{2005CBET..195....1M}, so is classed as an Fe\,{\sc ii}b nova. Most recently, the first spectrum taken of Nova SMC 2016 was consistent with it being a member of the He/N spectroscopic class \citep{2016ATel.9628....1W}. We note that Nova SMC 2016 has extensive pan-chromatic coverage (see e.g.\ \citealp{2016ATel.9635....1K,2016ATel.9688....1W,2016ATel.9733....1P,2016ATel.9810....1O}), which will be seen in forthcoming publication(s). The lack of early decline spectra of SMC novae prevents any conclusions being made on the proportion of novae that are Fe\,{\sc ii} novae or how Nova IC\,1613 2015 compares to the SMC population.

There are clearly significant changes in the wavelength dependence of the underlying continuum in the early stages of the eruption, which is apparent from the spectra, where the slope of the continuum changes from $f_{\lambda}\propto\lambda^{-2.42\pm0.08}$ 1.59 days after discovery to $f_{\lambda}\propto\lambda^{-1.48\pm0.08}$ 6.59 days after eruption. This is supported by the photometry where the nova shows a much redder colour 3.6 days after discovery compared to 1.6 days after discovery. In a nova eruption the {\it B}, {\it V} and {\it i}$^{\prime\prime}$ magnitudes are not greatly affected by line emission until the latter stages.

If we ignore the first two photometry points (where we have already established the continuum has changed dramatically over a short time), the {\it B}, {\it V}, and {\it i}$^{\prime}$ light curves are relatively well described by a power law. We find $f_{B({\mathrm{\lambda}})}\propto{t^{-1.22}}$, $f_{V({\mathrm{\lambda}})}\propto{t^{-1.28}}$ and $f_{i^{\prime}({\mathrm{\lambda}})}\propto{t^{-1.46}}$ (where {\it t} is days since discovery), which indicates the nova is becoming increasingly blue as the eruption evolves (as seen between spectra 2 to 5). However examining the ({\it B}$-${\it V}) colour evolution shows the picture is slightly more complex. The ({\it B}$-${\it V}) colour becomes lower until $m_V\sim20$ (around day 30$-$40) and then turns around and ({\it B}$-${\it V}) begins increasing (getting redder). This behaviour is seen in other novae (see for example Figure 39 in \citealp{2014ApJ...785...97H} and Figure 7 in \citealp{2016ApJ...833..149D}). While this seems in general agreement with the power-law of the final spectrum becoming shallower, we also must note that while {\it B} and {\it V} magnitudes are not as influenced by line emission early in the eruption, during the nebular phase, lines such as [O\,{\sc iii}] become increasingly dominant (and thus affect the broadband colours, but are removed from the power-law fitting in Section\,\ref{sec:sed}).

The X-ray upper limits in Table\,\ref{tab:swift} indicate that either the SSS phase had not started yet before day 330 or that the nova did not become visible in soft X-rays at all. If this were a M31 nova, then the fast $t_2$ and reasonably high expansion velocity would (consistently) predict a SSS turn-on time of about 50--90 days for a subset with similar properties \citep[cf.][]{2014A&A...563A...2H}. This is in agreement with the optical spectra indicating that by day 57 the nova had entered its nebular phase, where the ejecta would have become optically thin to X-rays. However, note that only a fraction of novae in the M31 reference sample showed SSS emission, which cannot be explained by observational coverage alone \citep{2011A&A...533A..52H,2014A&A...563A...2H}.

If we assume that by day 57 the nuclear burning in the residual hydrogen envelope (i.e. the part that was not ejected) had already extinguished, then we can estimate an upper limit on the mass of this envelope. Following the approach described in \citet[][see also the relevant references therein]{2014A&A...563A...2H} a SSS turn-off time of 57~days would correspond to a burned mass of 2.7\tpower{-7} M$_{\sun}$. Such relatively low masses are rare, but have been estimated for a few fast novae from the M31 sample \citep{2014A&A...563A...2H}. However, we stress the fact that for an individual object a large variety of factors, such as eruption geometry or inclination angle, can play a role in obscuring an existing SSS emission component.

Furthermore, there will of course be systematic differences between the nova samples of IC~1613 and M31, the latter of which we have learnt a great deal about nova population properties from. Different metallicities have been found to affect the average nova properties: see for instance the comparisons of M31 and LMC novae by \citet{1993A&A...271..175D,2013AJ....145..117S} and the theoretical models of \citep{2006ApJS..167...59H,2013ApJ...779...19K}. It remains unclear whether these systematics are large enough to affect the SSS phase of the nova significantly (e.g.\ to confine it to the narrow gap between days 160--260). Without further evidence we could only speculate on the specific causes for the SSS non detection, and we refrain from doing so.

\section{Summary and Conclusions}

We have presented detailed photometric and spectroscopic observations of the Nova IC\,1613 2015 eruption, from the early optically thick stage, through the early decline and nebular phases. This is the first detailed study of a nova residing in an irregular dwarf galaxy beyond the much closer MCs. Here we summarise our observations and conclusions:

\begin{enumerate}
\item We have undertaken a detailed observing campaign of Nova IC\,1613 2015, with ground-based photometry and spectroscopy led by the LT, with further observations from LCO. We also obtained UV photometry and X-ray observations with {\it Swift}.
\item The light curve shows a relatively smooth decline and the nova is classified as a fast nova, with $t_{2(V)}=13\pm2$ and $t_{3(V)}=26\pm2$\,days. The absolute peak magnitude of the nova is $M_V=-7.93\pm0.08$, which is typical for a classical nova.
\item The X-ray observations taken between 40--330\,days after discovery detected no SSS emission associated with the nova.
\item The spectra show that the nova is a member of the `hybrid' spectroscopic class, with it initially showing relatively strong Fe\,{\sc ii} lines and comparable N\,{\sc ii} lines. By the time it had declined by two magnitudes, the N\,{\sc ii}/N\,{\sc iii} features are significantly stronger than Fe\,{\sc ii}.
\item One of the more unusual features is a strong emission line peaking at $\sim$6162\,\AA. We interpret this as likely due to O\,{\sc i} (6157\,\AA), or possibly N~{\sc ii}.
\item The FWHM of the H$\alpha$ emission line is measured at around 1750\,km\,s$^{-1}$ and shows relatively little change over the course of the eruption.
\item The H$\alpha$/H$\beta$ ratio initially increases through the early decline spectra (due to self-absorption; peaking at $7.5\pm1.0$), before declining in the nebular spectrum. This implies the {\it z}$^{\prime}$-band light curve may be significantly influenced by a strong O\,{\sc i} 8446\,\AA\ emission line, which in turn is caused by Ly$\beta$ fluorescence.
\item We also obtained a nebular spectrum of Nova IC\,1613 2015, with [N\,{\sc ii}], [O\,{\sc i}], [O\,{\sc ii}] and [O\,{\sc iii}] all detected. This makes it one of the first novae beyond the MCs to be observed in the nebular phase.
\item The first spectrum taken around peak shows a steep blue continuum of $f_{\lambda}\propto\lambda^{-2.42\pm0.08}$, similar to that expected from optically thick free-free emission, but also consistent with photospheric (black-body) emission. The second spectrum, shows a dramatic change in the continuum to $f_{\lambda}\propto\lambda^{-1.48\pm0.08}$. A sudden change in the underlying continuum between the two epochs is supported by the photometry.
\item Despite the very close proximity of the nova to a stellar source, we find that this is most likely a chance alignment.
\end{enumerate}

To further our understanding of how nova eruptions depend on the underlying stellar population it is important we take the opportunity to study novae occurring in significantly different environments than can be found in the usual targets of M31 and our own Galaxy.

\section*{Acknowledgements}

We would like to thank the referee, Mike Shara, whose suggestions helped improve this paper. SCW acknowledges a visiting research fellowship at Liverpool John Moores University (LJMU). MH acknowledges the support of the Spanish Ministry of Economy and Competitiveness (MINECO) under the grant FDPI-2013-16933 as well as the support of the Generalitat de Catalunya/CERCA programme.  The Liverpool Telescope is operated on the island of La Palma by LJMU in the Spanish Observatorio del Roque de los Muchachos of the Instituto de Astrofisica de Canarias with financial support from the UK Science and Technology Facilities Council. This work makes use of observations from the LCO network. This work made use of data supplied by the UK {\it Swift} Science Data Centre at the University of Leicester. Based in part on data obtained from the ESO Science Archive Facility under request numbers scw233242 and scw233245.

%%%%%%%%%%%%%%%%%%%%%%%%%%%%%%%%%%%%%%%%%%%%%%%%%%

%%%%%%%%%%%%%%%%%%%% REFERENCES %%%%%%%%%%%%%%%%%%

% The best way to enter references is to use BibTeX:

%\bibliographystyle{mnras}
%\bibliography{mnras_template} % if your bibtex file is called example.bib

% Alternatively you could enter them by hand, like this:
% This method is tedious and prone to error if you have lots of references
% \begin{thebibliography}{99}
% \bibitem[\protect\citeauthoryear{Author}{2012}]{Author2012}
% Author A.~N., 2013, Journal of Improbable Astronomy, 1, 1
% \bibitem[\protect\citeauthoryear{Others}{2013}]{Others2013}
% Others S., 2012, Journal of Interesting Stuff, 17, 198
% \end{thebibliography}

%%%%%%%%%%%%%%%%%%%%%%%%%%%%%%%%%%%%%%%%%%%%%%%%%%

%%%%%%%%%%%%%%%%% APPENDICES %%%%%%%%%%%%%%%%%%%%%

\appendix

\section{NUV, Optical, and NIR Photometry of Nova IC\,1613 2015} \label{append}

\begin{table*}
\caption{Near-UV, optical, and near-IR photometry of Nova IC\,1613 2015. These data have not been corrected for any reddening.\label{tab:phot}}
\begin{tabular}{llllll}
\hline
Date [UT]& $t$ [days] & Telescope \& Instrument & Exposure [s] & Filter & Photometry \\
\hline
2016 Sep 20.993 &10.513 & LT IO:O &180 &{\it u$^{\prime}$} &$18.078\pm0.033$\\
2016 Sep 21.978 &11.498 & LT IO:O &120 &{\it u$^{\prime}$} &$17.954\pm0.038$\\
2016 Sep 25.046 &14.566 & LT IO:O &180 &{\it u$^{\prime}$} &$18.618\pm0.048$\\
2016 Sep 25.948 &15.468 & LT IO:O &180 &{\it u$^{\prime}$} &$18.653\pm0.064$\\
2016 Sep 27.036 &16.556 & LT IO:O &360 &{\it u$^{\prime}$} &$18.583\pm0.053$\\
2016 Sep 30.962 &20.482 & LT IO:O &360 &{\it u$^{\prime}$} &$18.894\pm0.044$\\
2016 Oct 06.009 &25.529 & LT IO:O &240 &{\it u$^{\prime}$} &$19.181\pm0.030$\\
2016 Oct 08.078 &27.598 & LT IO:O &360 &{\it u$^{\prime}$} &$19.353\pm0.029$\\
2016 Oct 09.014 &28.534 & LT IO:O &360 &{\it u$^{\prime}$} &$19.396\pm0.028$\\
2016 Oct 13.014 &32.534 & LT IO:O &360 &{\it u$^{\prime}$} &$19.527\pm0.031$\\
2016 Nov 03.995 &54.515 & LT IO:O &360 &{\it u$^{\prime}$} &$20.317\pm0.036$\\
2016 Nov 07.983 &58.503 & LT IO:O &360 &{\it u$^{\prime}$} &$20.421\pm0.040$\\
2016 Nov 15.967 &66.487 & LT IO:O &360 &{\it u$^{\prime}$} &$20.741\pm0.040$\\
2016 Dec 12.868 &93.388 & LT IO:O &360 &{\it u$^{\prime}$} &$21.554\pm0.062$\\
\hline
2016 Sep 12.087 &1.607 & LT IO:O &180 &{\it B} &$16.827\pm0.038$\\
2016 Sep 14.043 &3.563 & LT IO:O &60 &{\it B} &$17.171\pm0.038$\\
2016 Sep 17.114 &6.634 & LT IO:O &60 &{\it B} &$17.675\pm0.038$\\
2016 Sep 20.996 &10.516 & LT IO:O &180 &{\it B} &$18.252\pm0.038$\\
2016 Sep 21.981 &11.501 & LT IO:O &120 &{\it B} &$18.226\pm0.039$\\
2016 Sep 24.606 &14.126 & LCO 2\,m Spectral &360 &{\it B} &$18.927\pm0.068$\\
2016 Sep 25.049 &14.569 & LT IO:O &180 &{\it B} &$18.764\pm0.051$\\
2016 Sep 25.951 &15.471 & LT IO:O &180 &{\it B} &$18.925\pm0.047$\\
2016 Sep 27.057 &16.577 & LT IO:O &180 &{\it B} &$18.816\pm0.050$\\
2016 Sep 29.929 &19.449 & LT IO:O &180 &{\it B} &$18.932\pm0.055$\\
2016 Oct 3.576 &23.096 & LCO 2\,m Spectral &360 &{\it B} &$19.461\pm0.045$\\
2016 Oct 05.994 &25.514 & LT IO:O &180 &{\it B} &$19.430\pm0.039$\\
2016 Oct 08.062 &27.582 & LT IO:O &180 &{\it B} &$19.567\pm0.039$\\
2016 Oct 08.999 &28.519 & LT IO:O &180 &{\it B} &$19.579\pm0.039$\\
2016 Oct 10.920 &30.440 & LT IO:O &180 &{\it B} &$19.671\pm0.042$\\
2016 Oct 12.999 &32.519 & LT IO:O &180 &{\it B} &$19.803\pm0.040$\\
2016 Oct 18.969 &38.489 & LT IO:O &180 &{\it B} &$20.041\pm0.065$\\
2016 Oct 20.530 &40.050 & LCO 2\,m Spectral &360 &{\it B} &$20.154\pm0.073$\\
2016 Oct 30.038 &49.558 & LT IO:O &180 &{\it B} &$20.275\pm0.063$\\
2016 Nov 03.980 &54.500 & LT IO:O &180 &{\it B} &$20.362\pm0.043$\\
2016 Nov 07.969 &58.489 & LT IO:O &180 &{\it B} &$20.500\pm0.046$\\
2016 Nov 15.951 &66.471 & LT IO:O &120 &{\it B} &$20.698\pm0.052$\\
2016 Dec 12.845 &93.365 & LT IO:O &240 &{\it B} &$21.255\pm0.048$\\
\hline
2016 Sep 12.090 &1.610 & LT IO:O &180 &{\it V} &$16.651\pm 0.041$\\
2016 Sep 14.044 &3.564 & LT IO:O &60 &{\it V} &$16.885\pm0.041$\\
2016 Sep 17.115 &6.635 & LT IO:O &60 &{\it V} &$17.493\pm0.041$\\
2016 Sep 18.499 &8.020 & LCO 2\,m Spectral &360 &{\it V} &$17.953\pm0.058$\\
2016 Sep 20.978 &10.498 & LT IO:O &180 &{\it V} &$18.122\pm0.041$\\
2016 Sep 21.984 &11.504 & LT IO:O &180 &{\it V} &$18.118\pm0.037$\\
2016 Sep 24.611 &14.131 & LCO 2\,m Spectral &360 &{\it V} &$18.722\pm0.056$\\
2016 Sep 25.051 &14.571 & LT IO:O &180 &{\it V} &$18.716\pm0.045$\\
2016 Sep 25.954 &15.474 & LT IO:O &120 &{\it V} &$18.767\pm0.051$\\
2016 Sep 27.061 &16.581 & LT IO:O &180 &{\it V} &$18.749\pm0.047$\\
2016 Sep 29.934 &19.454 & LT IO:O &180 &{\it V} &$19.012\pm0.051$\\
2016 Sep 30.949 &20.469 & LT IO:O &180 &{\it V} &$19.078\pm0.045$\\
2016 Oct 03.580 &23.100 & LCO 2\,m Spectral &360 &{\it V} &$19.418\pm0.048$\\
2016 Oct 05.996 &25.516 & LT IO:O &180 &{\it V} &$19.457\pm0.042$\\
2016 Oct 09.001 &28.521 & LT IO:O &180 &{\it V} &$19.729\pm0.043$\\
2016 Oct 10.923 &30.443 & LT IO:O &180 &{\it V} &$19.791\pm0.044$\\
2016 Oct 13.002 &32.522 & LT IO:O &180 &{\it V} &$19.939\pm0.044$\\
2016 Oct 18.974 &38.494 & LT IO:O &120 &{\it V} &$20.077\pm0.050$\\
2016 Oct 20.536 &40.056 & LCO 2\,m Spectral &360 &{\it V} &$20.147\pm0.077$\\
2016 Oct 30.041 &49.561 & LT IO:O &180 &{\it V} &$20.343\pm0.101$\\
2016 Nov 03.983 &54.503 & LT IO:O &180 &{\it V} &$20.588\pm0.060$\\
2016 Nov 07.972 &58.492 & LT IO:O &180 &{\it V} &$20.485\pm0.079$\\
2016 Nov 15.954 &66.474 & LT IO:O &180 &{\it V} &$20.696\pm0.069$\\
2016 Dec 12.849 &93.369 & LT IO:O &300 &{\it V} &$21.089\pm0.052$\\
\hline
\end{tabular}
\end{table*}

\begin{table*}
\contcaption{Near-UV, optical, and near-IR photometry of Nova IC\,1613 2015. These data have not been corrected for reddening.\label{tab:phot_cont1}}
\begin{tabular}{llllll}
\hline
Date [UT]& $t$ [days] & Telescope \& Instrument & Exposure [s] & Filter & Photometry \\
\hline
2016 Sep 18.503 &8.023 & LCO 2\,m Spectral &180 &{\it r}$^{\prime}$ &$17.309\pm 0.015$\\
2016 Sep 21.002 &10.522 & LT IO:O &180 &{\it r}$^{\prime}$ &$17.550\pm 0.005$\\
2016 Sep 21.987 &11.507 & LT IO:O &120 &{\it r}$^{\prime}$ &$17.537\pm 0.006$\\
2016 Sep 24.616 &14.136 & LCO 2\,m Spectral &360 &{\it r}$^{\prime}$ &$17.967\pm 0.013$\\
2016 Sep 25.054 &14.574 & LT IO:O &180 &{\it r}$^{\prime}$ &$17.962\pm 0.009$\\
2016 Sep 25.957 &15.477 & LT IO:O &180 &{\it r}$^{\prime}$ &$18.032\pm 0.011$\\
2016 Sep 27.062 &16.582 & LT IO:O &180 &{\it r}$^{\prime}$ &$18.051\pm 0.012$\\
2016 Sep 29.938 &19.458 & LT IO:O &180 &{\it r}$^{\prime}$ &$18.277\pm 0.030$\\
2016 Sep 30.952 &20.472 & LT IO:O &180 &{\it r}$^{\prime}$ &$18.314\pm 0.010$\\
2016 Oct 03.585 &23.105 & LCO 2\,m Spectral &360 &{\it r}$^{\prime}$ &$18.371\pm 0.009$\\
2016 Oct 05.999 &25.519 & LT IO:O &60 &{\it r}$^{\prime}$ &$18.453\pm 0.009$\\
2016 Oct 08.068 &27.588 & LT IO:O &180 &{\it r}$^{\prime}$ &$18.532\pm 0.007$\\
2016 Oct 09.004 &28.524 & LT IO:O &180 &{\it r}$^{\prime}$ &$18.554\pm 0.006$\\
2016 Oct 10.925 &30.445 & LT IO:O &120 &{\it r}$^{\prime}$ &$18.579\pm 0.009$\\
2016 Oct 13.004 &32.524 & LT IO:O &180 &{\it r}$^{\prime}$ &$18.651\pm 0.007$\\
2016 Oct 18.976 &38.496 & LT IO:O &180 &{\it r}$^{\prime}$ &$18.845\pm 0.012$\\
2016 Oct 20.541 &40.061 & LCO 2\,m Spectral &360 &{\it r}$^{\prime}$ &$18.892\pm 0.015$\\
2016 Oct 30.044 &49.564 & LT IO:O &180 &{\it r}$^{\prime}$ &$19.269\pm 0.017$\\
2016 Nov 03.985 &54.505 & LT IO:O &60 &{\it r}$^{\prime}$ &$19.487\pm 0.021$\\
2016 Nov 07.974 &58.494 & LT IO:O &180 &{\it r}$^{\prime}$ &$19.614\pm 0.016$\\
2016 Nov 15.957 &66.477 & LT IO:O &180 &{\it r}$^{\prime}$ &$20.003\pm 0.016$\\
2016 Dec 12.853 &93.373 & LT IO:O &240 &{\it r}$^{\prime}$ &$20.881\pm 0.026$\\
\hline
2016 Sep 12.094 &1.614 & LT IO:O &360 &{\it i}$^{\prime}$ &$16.891\pm 0.007$\\
2016 Sep 14.045 &3.565 & LT IO:O &60 &{\it i}$^{\prime}$ &$16.762\pm 0.008$\\
2016 Sep 17.116 &6.636 & LT IO:O &60 &{\it i}$^{\prime}$ &$17.185\pm 0.008$\\
2016 Sep 18.506 &8.026 & LCO 2\,m Spectral &180 &{\it i}$^{\prime}$ &$17.462\pm 0.018$\\
2016 Sep 21.005 &10.525 & LT IO:O &180 &{\it i}$^{\prime}$ &$17.830\pm 0.007$\\
2016 Sep 21.990 &11.510 & LT IO:O &180 &{\it i}$^{\prime}$ &$17.877\pm 0.007$\\
2016 Sep 24.621 &14.141 & LCO 2\,m Spectral &360 &{\it i}$^{\prime}$ &$18.427\pm 0.021$\\
2016 Sep 25.057 &14.577 & LT IO:O &180 &{\it i}$^{\prime}$ &$18.425\pm 0.011$\\
2016 Sep 25.960 &15.480 & LT IO:O &180 &{\it i}$^{\prime}$ &$18.509\pm 0.015$\\
2016 Sep 27.065 &16.585 & LT IO:O &180 &{\it i}$^{\prime}$ &$18.553\pm 0.019$\\
2016 Sep 29.941 &19.461 & LT IO:O &180 &{\it i}$^{\prime}$ &$18.814\pm0.021$\\
2016 Sep 30.955 &20.475 & LT IO:O &180 &{\it i}$^{\prime}$ &$18.899\pm0.017$\\
2016 Oct 03.590 &23.110 & LCO 2\,m Spectral &360 &{\it i}$^{\prime}$ &$19.144\pm0.019$\\
2016 Oct 06.002 &25.522 & LT IO:O &180 &{\it i}$^{\prime}$ &$19.243\pm0.012$\\
2016 Oct 08.071 &27.591 & LT IO:O &180 &{\it i}$^{\prime}$ &$19.400\pm0.013$\\
2016 Oct 09.007 &28.527 & LT IO:O &180 &{\it i}$^{\prime}$ &$19.444\pm0.013$\\
2016 Oct 10.928 &30.448 & LT IO:O &180 &{\it i}$^{\prime}$ &$19.529\pm0.019$\\
2016 Oct 13.007 &32.527 & LT IO:O &180 &{\it i}$^{\prime}$ &$19.642\pm0.016$\\
2016 Oct 20.544 &40.064 & LCO 2\,m Spectral &360 &{\it i}$^{\prime}$ &$20.013\pm0.051$\\
2016 Oct 30.047 &49.567 & LT IO:O &180 &{\it i}$^{\prime}$ &$20.415\pm0.047$\\
2016 Nov 03.988 &54.508 & LT IO:O &180 &{\it i}$^{\prime}$ &$20.548\pm0.038$\\
2016 Nov 07.977 &58.497 & LT IO:O &180 &{\it i}$^{\prime}$ &$20.596\pm0.049$\\
2016 Nov 15.960 &66.480 & LT IO:O &180 &{\it i}$^{\prime}$ &$20.832\pm0.049$\\
2016 Dec 12.866 &93.386 & LT IO:O &300 &{\it i}$^{\prime}$ &$21.248\pm0.138$\\
\hline
2016 Sep 21.008 &10.528 & LT IO:O &180 &{\it z}$^{\prime}$ &$17.431\pm 0.015$\\
2016 Sep 21.993 &11.513 & LT IO:O &120 &{\it z}$^{\prime}$ &$17.443\pm 0.015$\\
2016 Sep 25.060 &14.580 & LT IO:O &180 &{\it z}$^{\prime}$ &$17.820\pm 0.018$\\
2016 Sep 25.963 &15.483 & LT IO:O &180 &{\it z}$^{\prime}$ &$17.879\pm 0.027$\\
2016 Sep 27.068 &16.588 & LT IO:O &180 &{\it z}$^{\prime}$ &$17.904\pm 0.022$\\
2016 Sep 27.940 &17.460 & LT IO:O &180 &{\it z}$^{\prime}$ &$17.990\pm 0.039$\\
2016 Sep 29.944 &19.464 & LT IO:O &180 &{\it z}$^{\prime}$ &$18.065\pm0.023$\\
2016 Sep 30.958 &20.478 & LT IO:O &180 &{\it z}$^{\prime}$ &$18.078\pm0.019$\\
2016 Oct 06.005 &25.525 & LT IO:O &120 &{\it z}$^{\prime}$ &$18.172\pm0.018$\\
2016 Oct 08.074 &27.594 & LT IO:O &120 &{\it z}$^{\prime}$ &$18.322\pm0.019$\\
2016 Oct 09.010 &28.530 & LT IO:O &180 &{\it z}$^{\prime}$ &$18.337\pm0.016$\\
2016 Oct 13.010 &32.530 & LT IO:O &180 &{\it z}$^{\prime}$ &$18.556\pm0.018$\\
2016 Oct 18.981 &38.501 & LT IO:O &180 &{\it z}$^{\prime}$ &$18.950\pm0.050$\\
2016 Oct 30.050 &49.570 & LT IO:O &120 &{\it z}$^{\prime}$ &$19.689\pm0.089$\\
2016 Nov 03.992 &54.512 & LT IO:O &180 &{\it z}$^{\prime}$ &$19.940\pm0.085$\\
\hline
\end{tabular}
\end{table*}

\begin{table*}
\contcaption{Near-UV, optical, and near-IR photometry of Nova IC\,1613 2015. These data have not been corrected for reddening.\label{tab:phot_cont2}}
\begin{tabular}{llllll}
\hline
Date [UT]& $t$ [days] & Telescope \& Instrument & Exposure [s] & Filter & Photometry \\
\hline
2016 Nov 07.980 &58.500 & LT IO:O &180 &{\it z}$^{\prime}$ &$20.057\pm0.106$\\
2016 Nov 15.963 &66.483 & LT IO:O &180 &{\it z}$^{\prime}$ &$20.536\pm0.084$\\
2016 Dec 12.863 &93.383 & LT IO:O &300 &{\it z}$^{\prime}$ &$21.366\pm0.120$\\
\hline
2016 Sep 19.191 &8.711 & LT IO:I &538 &{\it H} &$16.598\pm0.103$\\
2016 Sep 21.043 &10.563 & LT IO:I &538 &{\it H} &$16.991\pm0.117$\\
2016 Sep 25.066 &14.586 & LT IO:I &538 &{\it H} &$17.678\pm0.102$\\
2016 Oct 08.965 &28.485 & LT IO:I &538 &{\it H} &$18.772\pm0.143$\\	
\hline
\end{tabular}
\end{table*}

%%%%%%%%%%%%%%%%%%%%%%%%%%%%%%%%%%%%%%%%%%%%%%%%%%

% Don't change these lines
\bsp	% typesetting comment
\label{lastpage}
\end{document}